\documentclass[12pt]{article} 
\usepackage[margin=1in]{geometry}
\usepackage{tabularx, booktabs}
\usepackage{varwidth}
\usepackage{adjustbox}
\newcolumntype{Y}{>{\centering\arraybackslash}X}

\usepackage{datetime}           
\usdate                         

\usepackage[labelfont=bf,textfont=sl,tableposition=top]{caption}
\usepackage[pdftex,bookmarks,colorlinks]{hyperref}
\usepackage{amssymb,amsmath,amsthm,amsfonts,dsfont}
\usepackage{url,graphicx,multicol}
\usepackage{verbatim,courier}
\usepackage{natbib}

\usepackage{array}
\usepackage{tabularx}
\usepackage{multirow}
\usepackage{color}
\usepackage{authblk}
\usepackage{longtable}
\usepackage{dsfont} 

\usepackage{algorithm}
\usepackage{algorithmic}

\usepackage{mathtools}

\usepackage{epstopdf}
\definecolor{navy}{rgb}{0,0,0.502}
\definecolor{greenb}{rgb}{0,0.7,0}
\hypersetup{colorlinks,%
citecolor=navy,%
filecolor=green,%
linkcolor=black,%
urlcolor=blue,%
}

\usepackage[autostyle]{csquotes}  
\usepackage{soul}
\usepackage{booktabs}

\newcolumntype{R}{>{\raggedleft\arraybackslash}p{6cm}}

\newcommand\NECO{\mbox{NECO}}
\newcommand\NECOF{\mbox{NECOF}}
\newcommand\NECOFest{\widehat{\mbox{NECOF}}}
\makeatletter
\newcommand*{\indep}{%
  \mathbin{%
    \mathpalette{\@indep}{}%
  }%
}
\newcommand*{\nindep}{%
  \mathbin{
    \mathpalette{\@indep}{\not}
  }%
}
\newcommand*{\@indep}[2]{%
  \sbox0{$#1\perp\m@th$}
  \sbox2{$#1=$}
  \sbox4{$#1\vcenter{}$}
  \rlap{\copy0}
  \dimen@=\dimexpr\ht2-\ht4-.2pt\relax
  \kern\dimen@
  {#2}%
  \kern\dimen@
  \copy0 
} 
\makeatother

\usepackage{marginnote,datetime,url,enumitem,subfigure,rotating}

\makeatletter
\def\ThisWidth#1{%
  \@tempdima\dimexpr(\textwidth-#1)/2\relax
  \edef\FRleftmargin{\noexpand\hspace*{\the\@tempdima}}%
  \edef\FRrightmargin{\noexpand\hspace*{\the\@tempdima}}%
  \ignorespaces}
\makeatother

\usepackage{tikz}
\usetikzlibrary{arrows,shapes}
\usetikzlibrary{shapes,decorations,arrows,calc,arrows.meta,fit,positioning}
\tikzset{
-Latex,auto,node distance =1 cm and 1 cm,semithick,
intervention/.style ={rectangle, draw=black},
random/.style ={circle, draw=black},
parameter/.style ={diamond, draw=black},
}
\begin{document}


\date{\today}

\title{Using Network-based Causal Inference to Detect the Sources of Contagion in the Currency Market
}
  
\author{Katerina Rigana\thanks{Swiss Finance Institute, Università della Svizzera Italiana (USI), Lugano, Switzerland. E-mail: katerina.rigana@usi.ch.}, Ernst C. Wit, Samantha Cook\\
} 
  
\maketitle

\begin{abstract}
%
%
Contagion is an extremely important topic in finance. Contagion is at the core of most major financial crises, in particular the 2008 financial crisis. Although various approaches to quantifying contagion have been proposed, many of them lack a causal interpretation. We will present a new measure for contagion among individual currencies within the Foreign exchange market and show how the paths of contagion work within the Forex using causal inference. This approach will allow us to pinpoint sources of contagion and to find which currencies offer good options for diversification and which are more susceptible to systemic risk, ultimately resulting in feedback on the level of global systemic risk.
\end{abstract}




\newpage

\newpage

\section{Introduction }

The global economy continues to be more and more integrated and interconnected. This presents enormous opportunities for growth and innovation, as well as new types of risks. In order to see how resilient financial networks are to contagion, financial regulators need to understand how contagion propagates and where the sources of contagion reside \citep{stiglitz2010contagion}. 
There are various ways to define financial contagion.
The first descriptions of contagion are predominantly in terms of what today we would call behavioural finance --- in terms of sentiments, emotions, behavioural biases and crowd effects \citep{hansen2021financial}. 
Modern research into contagion within the academic literature began to appear in the wake of the 1987 market crash and exploded during the 1995 Mexican crisis, the 1997 Asian financial crisis and the related 1998 Russian financial crisis that followed. The main goal was to explain how a series of potentially local issues could spread from country to country, creating a financial crisis with global repercussions \citep{edwards2000contagion, claessens2013international}. These events caught many economists off guard and so most papers during this period concentrate on explaining \citep{edwards2000contagion, calvo1996capital}, or explaining away \citep{forbes2002no, collins2003contagion, karolyi2004does}, this new market behaviour. Most of the definitions of contagion during the period define it as  shocks or correlations that are unexplained or unexpected or significantly higher than usual, in contrast to expected and explainable changes and correlations, called spillovers and interdependence \citep{rigobon2019contagion}. 
The 2007-2008 financial crisis rekindled interest into analysing contagion. This time around there has been less research into whether contagion exists and more into how to estimate and model it properly to be ready for future crises. During this period we see the appearance of network theory applications that set the tone for future research \citep{nier2007network, gai2011complexity, elliott2014financial, glasserman2015likely, cook2016network, battiston2018financial}.

There have been various approaches to contagion estimation: the copula approach \citep{rodriguez2007measuring, wen2012measuring};
the vector moving average and variance decomposition, related to the Impulse response modelling \citep{diebold2014network, barigozzi2017network, barigozzi2020time};
and new approaches of GARCH estimation to structural models \citep{dahlhaus2003vcausality, eichler2007granger, giudici2018corisk, avdjiev2019measuring}. Each of these methods has its own advantages but they lack an intervention-based causality interpretation. Traditionally when it comes to causality economists have relied on the Granger Causality concept \citep{stavroglou2017causality}, but this concept merely relies on temporal correlations rather then structural causation \citep{sugihara2012detecting}.

We will analyse contagion through changes in price due to factors spreading from instrument to instrument that cannot be explained by individual trends. This type of contagion can be interpreted as \enquote{pure contagion} defined by \cite{van2001sources}. Combining causal networks with a structural Vector Autoregression (VAR) model, we are able to extract from the log returns of the exchange rates which part of a change in the value of a currency is caused by the idiosyncratic characteristics of the currency and which part is caused by contemporaneous contagion effects from other currencies. The structural VAR part of our approach is an extension of \cite{giudici2018corisk}. Similar VAR approaches have been used to analyse  systemic risk within the foreign exchange market; none, however, look at the causal direction of such effects. 
In order to analyse the contagion paths in the foreign exchange market, we will recreate directed partial correlation based networks using the causality concepts from \cite{pearl2009causality} extended into practical methods by \cite{spirtes1991algorithm} and \cite{colombo2014order}. 

The rest of this paper is organized as follows. In Section~\ref{theory}, we first explain basic concepts of network theory and causal inference and then describe how we estimate contagion using Causal Graphical models within a structural VAR model. In Section~\ref{empiricalresults} we analyse contagion within a subset of currencies on the Forex during the years 2000-2021; and we conclude in Section~\ref{conclusion} by summarising the main advantages of this approach to measuring contagion in finance, while noting the remaining challenges.

\section{A Structural Equation Model for Contagion} \label{theory}

Our measure of contagion within a network relies on causal graphical model theory. In this section we will introduce the main underlying concepts and notation. 
A network is a collection or system of interconnected objects such as things, people or groups of people (like institutions or countries, for example). In mathematics these interconnections are represented as graphs, defined as a set of nodes or vertices that are connected by links or edges. 
In this work we will analyse directed networks by estimating causal links that are not directly observed between financial instruments or assets. Such networks are called causal networks, a specific type of a graphical model (also often called Bayesian or belief networks). These networks incorporate probabilistic relationships between the nodes in the form of a Directed Acyclic Graph or DAG \citep{Howard1981,Pearl1988}. 


Bayesian networks can be analysed using several different types of algorithms. There are the so-called constraint-based algorithms  that look for conditional independence \citep{colombo2014order} and the main algorithm of this group, the PC-algorithm \citep{spirtes1991algorithm}, is the one we will use in Section \ref{source}. The other type of algorithms are the so-called score-based algorithms, which maximise a causal objective function. 
A comparison between these methods in terms of speed and accuracy can be found in \cite{scutari2018learns}. Importantly, they find that \enquote{constraint-based algorithms are more accurate than score-based algorithms for small sample sizes.} 





\subsection{Causal Graphical Models} \label{explain}
Here we present the mathematics underlying causal graphical models. We consider $N$ assets that we want to analyse and their log returns are represented as the multivariate random variable $X = (X_{1}, \dots, X_{N})$.
%
Each of these N assets will represent a node in our network. We can analyse $X$ as a directed graphical model (GM) represented by a causal DAG. The arrows in the network $G_X$ are to be interpreted in terms of conditional independence (CI) with the additional causal interpretation.

%
%



A DAG $G_X$ is causal for a probability distribution $f(x)$ if $f(x)$ recursively factorizes with respect to $G_X$ and the following intervention formula holds for all subsets $A$ in $V$, the set of all nodes:
\begin{equation}\label{intervention}
	\forall A \subseteq V \quad f(x \Vert x^*_A) = \prod_{\alpha \in V\setminus A} f(x_{\alpha} \vert X_{pa(\alpha)} = x_{pa(\alpha)} ) \Bigr\rvert_{X_A \, = \, x^*_A} 	
\end{equation}\medskip
where $f(x \Vert x^*_A)=f(x \vert x_A \leftarrow x_{A}^{*})$ is the distribution of X under the knowledge that a subset $A$ of the nodes $V$ in the network have been imposed a value $x^{*}_{A}$. An example of this could be when a central bank decides to fix its exchange rate with respect to another currency and we want to see the impact on the other currencies in the market. The conditional distributions $f(x_{\alpha} \vert X_{pa(\alpha)})$ are assumed stable under interventions that do not involve $x_{\alpha}$ --- hence the condition that $\alpha$ represent nodes in $V$ but not in in the intervention subset $A$. This conditioning by intervention allows for much more specific causal interpretation than the conditional distribution. The random variable of interest $X$ is a causal GM if it is a directed GM, as described above, such that the intervention factorization in Equation \ref{intervention} holds. This definition is often referred to as the Lauritzen's causal graph with interventions by replacement \citep{lauritzen2001causal} or as Pearl's do-intervention \citep{pearl2009causality}.

\subsection{Defining Contagion} \label{measure}

In this section we will present a network-based measure for contagion. It is based on the causal graph models presented in the previous section, combined with an autoregressive approach to contagion as described in \cite{giudici2018corisk}. \cite{giudici2018corisk} introduce a structural vector autoregression (VAR) to analyse the contagion impact on the the cost of insuring public debt during the European sovereign debt crisis.


%
%
%
%
%
%
We propose a structural equation model for the evolution of the log returns $X_{it}$ on a financial asset $i$ at time $t$ through a structural VAR consisting of an autoregressive part $AR(i,t)$ and a network contagion part that we will call $NECO(i,t)$:

\begin{equation} \label{commented}
\underbrace{X_{i,t}}_{\mbox{LogReturn on Asset i}}
 \leftarrow 
 \alpha_{0,i}+ \overbrace{ \sum_{\ell=1}^{L} \alpha^{\ell}_{i}X_{i,t-l}}^{\mbox{AR(i,t)}}
+ \overbrace{\sum_{j\in pa(i)}\beta_{ji}\underbrace{X_{j,t}}_{\mbox{Assets that affect asset i}}}^{\mbox{Contagion=\mbox{NECO}(i,t)}}+ \underbrace{\varepsilon_{i,t}}_{\mbox{Noise}}
\end{equation}
where $\ell$ are the number of considered lags for the autoregressive part,  $\alpha^{\ell}_{i}$ are the autoregressive coefficients at lag $\ell$ for asset $i$ and $\beta_{ji}$ the causal effects of asset $j$ on asset $i$ or $X_{j}$ on $X_{i}$. In the causal literature, the causal effects are defined as the partial derivative of the expected log-return
$\frac{d}{dx_{j}}E(X_{i,t}\vert X_{-i,t},X_{i,t-\ell})$
which in our case is equal to $\beta_{ji}$. Just like we do for the autoregressive part, lags can be added to the $\NECO$ part. We leave this extension for future research. 
The arrow in Equation \ref{commented} is to be interpreted in a generative manner, such as in a structural equation model \citep{bollen2013eight}. The right-hand side is the driving force behind the value of $X_{it}$.
%
Equation \ref{commented} can be summarised as:
\begin{equation}\label{regression}
X_{it} = \alpha_{0,i}+\mbox{AR}_{it} + \mbox{NECO}_{it}+\varepsilon_{it}
\end{equation}
where the $NECO_{it}$ measures the totality of the contagion effect on the market. As a measure of the impact of contagion on the price of an individual asset, we propose the Network Contagion Factor ($\mbox{NECOF}$), which is computed as follows:

\begin{equation} \label{necofdef}
\mbox{NECOF}(i) = \frac{{\sigma}_{i,NC}^{2}}{\sigma_{i,AR}^{2}+\sigma_{i,NC}^{2}+\sigma_{i}^{2}} = 1 \, - \, \frac{\sigma_{i,AR}^{2}+\sigma_{i}^{2}}{\sigma_{i,AR}^{2}+\sigma_{i,NC}^{2}+\sigma_{i}^{2}}
\end{equation}
where $\sigma_{i,AR}^{2}=V(AR_{it})$, 
$\sigma_{i,NC}^{2}=V(\mbox{NECO}_{it})$ and $\sigma_{i}^{2}=V(\varepsilon_{it})$, under the assumption that $\mbox{AR}_{it}$, $\mbox{NECO}_{it}$ and $\varepsilon_{it}$ are independent. This independence assumption is realistic, given that the considered contagion is assumed to be instantaneous and therefore by definition isolated from the idiosyncratic effects $\mbox{AR}_{it}$.



The $\mbox{NECOF}$ is expressed in percentages and shows the impact of contagion on the return of the asset $i$. A $\mbox{NECOF}$ of 0\% would mean that contagion has no impact on the considered asset. On the opposite end of the spectrum, a $\mbox{NECOF}$ of 100\% would indicate that the impact of contagion for the given asset is absolute. The $\mbox{NECOF}$ measure alone is very useful to identify which assets are at higher risk of outside influence - information that can be useful for investment and diversification strategies alike. We call the $\mbox{NECO}$ a network-based measure because it depends on the underlying causal graph.



\subsection{Identifying Contagion Paths} \label{source}

%
 
This section shows how the causal networks and causal $\NECO$ coefficients $\beta_{ji}$ in Equation \ref{commented} are estimated. Once we estimate the $\NECO$ coefficients and the $\NECOF$ we can recover the contagion factor for each instrument. In order to estimate the causal coefficients correctly, we first need to establish the causal structure, which means finding all the causal parents for every considered asset $i$ in our graph.

We estimate the causal structure using a more robust version of the standard PC-algorithm from \cite{spirtes1991algorithm} called the PC-stable algorithm \citep{colombo2014order}\footnote{We perform the PC-Algorithm using the pcalg package in R as described in \cite{kalisch2020overview} }. The PC-Algorithm uses two steps in order to find the sources of contagion, which are summarised in Appendix \ref{SecPC}. We add a third step to estimate the size of the contagion effects, $\beta_{ji}$, by performing a series of linear regressions on Equation \ref{necofdef}. Given the set of parents for each asset, the non-zero $\beta_{ji}$ are to be estimated from the obtained DAG. If at the end of Step 2 we achieve a Completed Partially Directed Acyclic Graph (CPDAG), a subset of Markov equivalent DAGs that can explain our data, we will also find a multiset of possible $\mbox{NECO}$ estimates
$\hat{\beta}_{ij}$. We can combine these into a range estimator as in \cite{maathuis2009estimating}.
Once we have estimated the $\mbox{NECO}$ we can estimate the $\mbox{NECOF}$ of interest for each instrument in our model using the following equation:
\begin{equation} \label{necofSS}
\widehat{\mbox{NECOF}}_{i}=1-\frac{RSS_{NECO} (i) }{SS(i)}
\end{equation}
that we obtain by applying the Type II\footnote{Similar to Type I but not dependent on the order of entry of terms into the model \citep{yandell1997practical}.} Sums of Squares to Equation \ref{regression} such that:
\begin{equation}
RSS_{CR} (i) = \sum_{t}  \left[ X_{i,t} - \widehat{\mbox{NECO}}(i,t)\right]  ^{2}
\mbox{ and  }
SS(i)= \sum_{t} \left[ X_{i,t} - \mu_{i,t} \right] ^{2}
\end{equation}

%
%
%
%
\subsection{Community detection} \label{clustering}

Using the contagion paths established in the previous section we can identify communities of financial instruments. These communities are groups of nodes that are more connected among themselves within the group then with the other groups. These groups are called communities, clusters or modules. Communities can be seen as sub-graphs that have specific properties not shared by the whole network and this allows for a next-level analysis of the network, moving from a single node to a more meaningful structure. These communities can be also seen as meta-nodes when representing and analysing very large networks, where considering and plotting each singular node would not be practically feasible.

There are many different algorithms to identify communities within a network; for a comparative study see \cite{lancichinetti2009community}. We will be using the Louvain algorithm from \cite{blondel2008fast} to establish communities among the nodes, because it a benchmark among the clustering algorithm thanks to its robust and efficient results, making the results easier to compare with other studies. 


\subsection{Creating Dynamic Contagion Maps} \label{maps}

In the previous sections we defined a measurement of the contagion and the sources of this contagion, assuming that the $\NECO$ and its coefficients would remain constant for the entire time period in consideration. In this section we will add a dynamic component, and in doing so not only allow for the $\mbox{AR}(i,t)$ and $\mbox{NECO}(i,t)$ in Equation \ref{commented}  to change with time but also for the whole causal structure of our contagion graph to change with time. A dynamic version of (\ref{commented}) is written as,
\begin{equation} \label{time}
X_{it}\leftarrow \alpha_{0,i}+\sum_{l=1}^{L} \alpha_{l,i}^{t}X_{i,t-l}+\sum_{\forall j\in pa(i)}\beta_{ji}^{t}X_{j,t}+\varepsilon_{i,t}
\end{equation}
describing a dynamic causal graphical model. From an inferential point of view, we will estimate the coefficients in a piecewise-constant way. At each timepoint $t$ we evaluate a new DAG and the associated $\NECOFest$ estimates, creating a sequence of contagion maps. The contagion effect is considered to be contemporaneous within the considered window of time $[t-1,t]$. The length of the window will vary with the use case and depends on the data being analysed, what kind of short or long term trends are associated and the purpose of the study.
\section{Empirical Findings} \label{empiricalresults}

The Forex is an important financial market, trading \$6.6 trillion per day \citep{bank2019triennial}. Given that the most traded exchange rates are those over the USD we consider the interaction of 23 exchange rates over the USD for the years 2000-2021 as published daily by the Federal Reserve of New York. 
This allows us to evaluate the networks among highly traded currencies by expressing their value in terms of the US Dollar, the most liquid of all currencies, based on reliable historic data. Alternative approaches include using exchange rates based on the special drawing right (SDR) as in \cite{wang2012similarity}. SDR  reflects the price of a basket of five major currencies and is periodically rebalance and published on a daily basis by the International Monetary Fund (IMF). Another option for the base currency is to choose a currency that is of lesser importance, but still not completely illiquid. An example of this approach can be found in \cite{keskin2011topology} who use the Turkish Lira as a base. A comparison of different currencies being used as the base currency can be found in \cite{kwapien2009analysis}. One final approach taken by \cite{fenn2012dynamical} is that of ignoring the base currency issue altogether and using each exchange rate as a separate financial asset.
\subsection{Foreign Exchange Rates 2000-2021} \label{data}

We consider the log returns on the spot exchange rates. The distribution of log-returns on the exchange rates is often assumed normal, but as with most financial assets there is the presence of fat tails. \cite{johnston1999statistical} analyse this problem, without finding any better alternative that would hold for every currency and time frame. 
Some studies even find that trading strategies based on the assumption of log-normality do in fact maximise profit \citep{sarpong2019estimating}. 
The presence of fat tails will cause the significance level for the individual conditional independence tests within the PC-Algorithm to be empirical slightly higher than the nominal value.

\begin{figure}[tb]
\includegraphics[width=\textwidth]{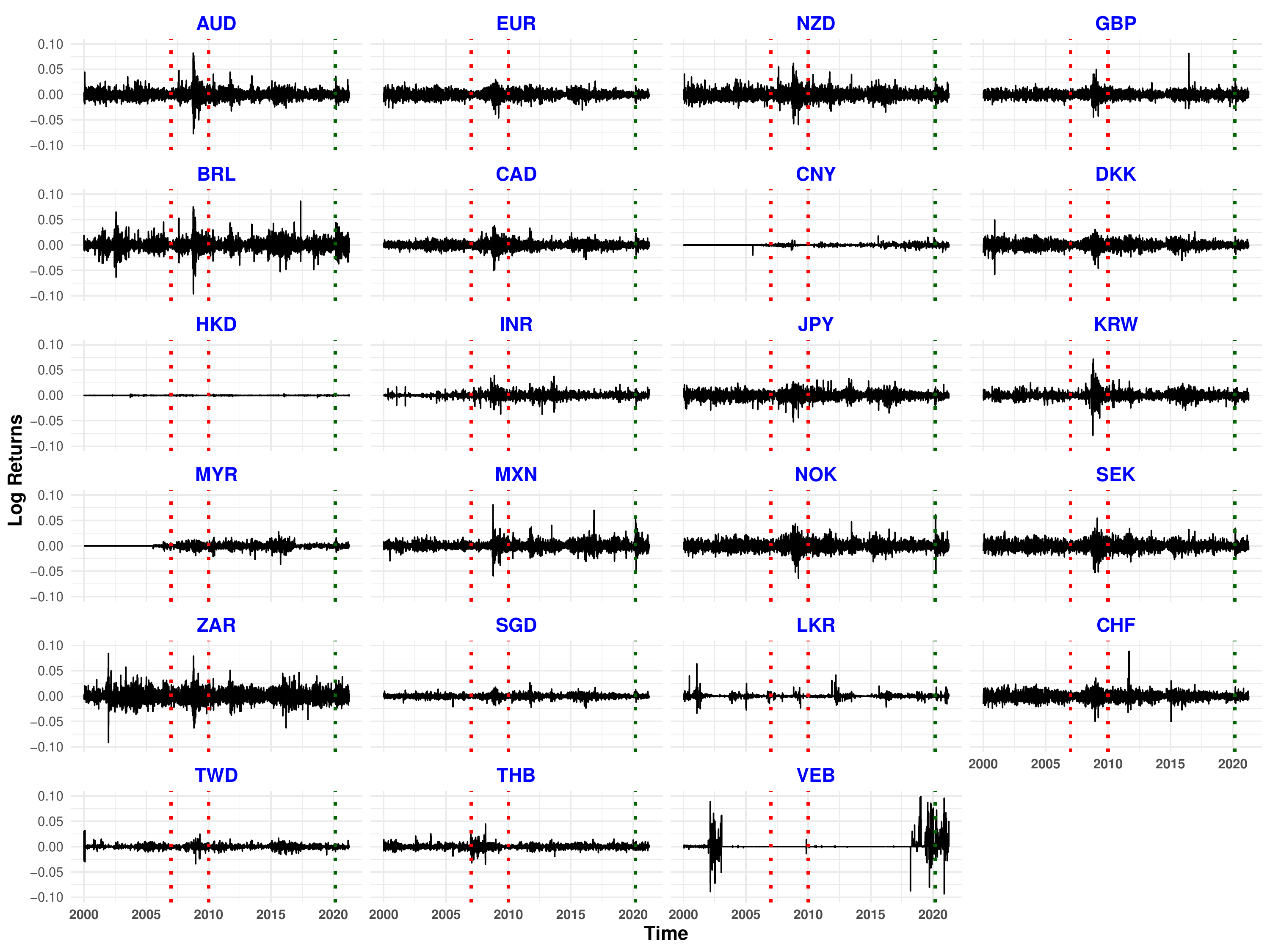}
\captionof{figure}[Log returns for each currency.]{Log returns for each currency, in red the financial crisis years from 2007 to 2009 and the green line shows the beginning of the Covid Recession.} 
\label{fig:logreturns}
\end{figure}

Figure \ref{fig:logreturns} shows the series of 23 log-returns for our time frame, from January 2000 until April 2021. The two red vertical lines in each chart delimiting the 2007-2009 period show that all of the exchange rates have been impacted by the Financial Crisis in some way; even highly managed currencies like the Venezuelan Bolívar (VEB) or the  Chinese Yuan Renminbi (CNY) show some turbulence during this period. The volatility increases for all currencies during the period 2007-2009. Flat periods where the exchange rate over the USD does not change at all reflect periods where the currency was pegged to the USD. This is the case for the CNY and the Malaysian Ringgit (MYR) in the early 2000s for example. China switched from a fixed exchange rate to a less restricted regime in July 2005, with Malaysia following suit.This has resulted in the value of those currencies getting closer to their perceived market value - for the MYR an appreciation and for the CNY a depreciation. The Chinese government is only slowly allowing more flexibility of the exchange rate \citep{chinapost20100619}, but it remains a highly influential rate and is the first emerging market currency to be held as a reserve by the International Monetary Fund (IMF). The Venezuelan Bolívar (VEB)  also shows an unusual history of returns. This dynamic reflects periods of hyperinflation and the subsequent government sanctioned devaluation of the currency.

\begin{figure}[tb]
\begin{tabular}{cc}
\includegraphics[width=0.4\textwidth]{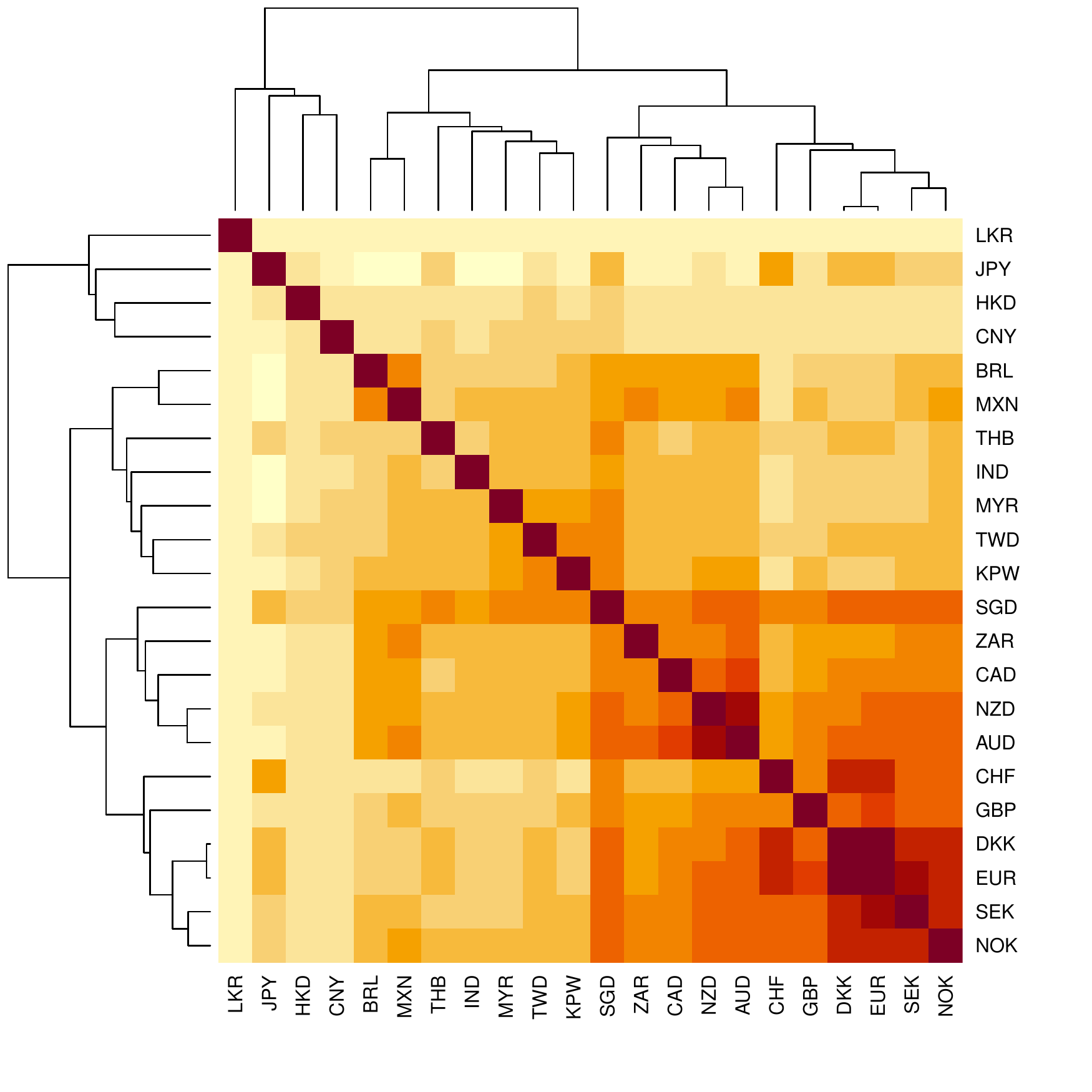}
&\includegraphics[width=0.6\textwidth]{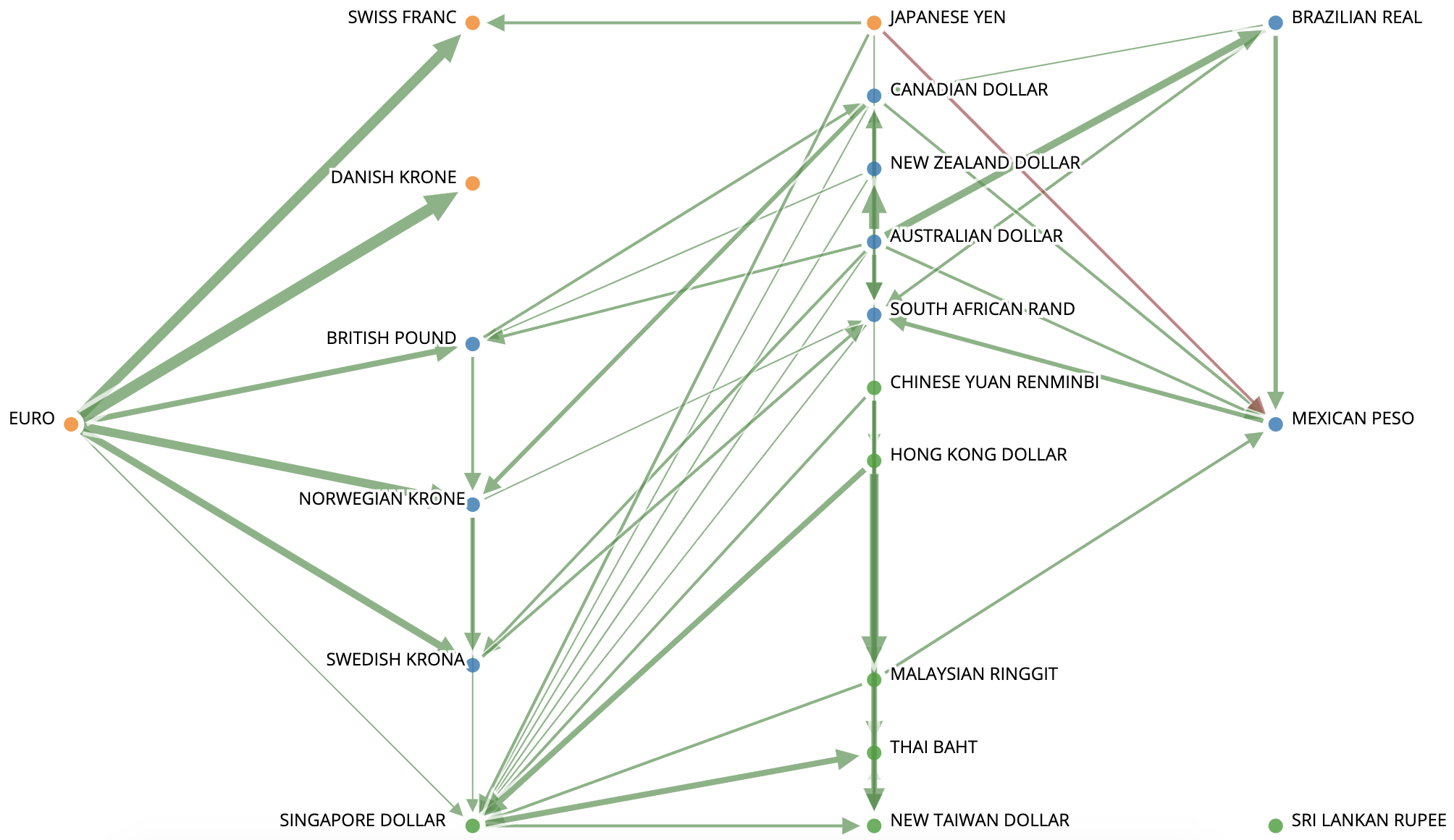}\\
(a)&(b)
\end{tabular}
\captionof{figure}[(a) Correlation heatmap for the time period 2000-2021; (b) Causal network showing the interconnections and contagion paths within the 23 foreign exchange rates for years 2000-2021.]{(a) Correlation heatmap for the time period 2000-2021; (b) Causal network showing the interconnections and contagion paths within the 23 foreign exchange rates for years 2000-2021. Note that colors represent Louvain clusters; green (red) arrows indicate a positive (negative) contagion coefficient of the corresponding causal effect; the width of the arrows reflect the strength of the causal effect.}
\label{fig:heatmap}
\end{figure}

Figure \ref{fig:heatmap}a shows the correlation heatmap between the considered currencies. This is what a static approach using correlation networks can reveal. The heat map can detect the outlines of the main clusters present on the Forex. The clusters include a European cluster (EUR, NOK, SEK, DKK, GDP, CHF), the Commonwealth cluster (AUD, NZD, CAD, ZAR, SGP), a small cluster of emerging economies (BRL and MXN) and then a somewhat sparse geographically based cluster of the Asian currencies. Applying the models from section \ref{explain} we will be able look more deeply and more detail into the dynamics within the Forex of these currencies.
 
\subsection{Contagion in the Currency Market}

This section will present results of applying the methods from Section \ref{theory} to the Forex returns data described above.
Figure \ref{fig:heatmap}(b) shows the causal network created from the complete set of returns data (2000 - 2021). The nodes are the 23 currencies in our dataset and the links represent the causal effects of contagion from one currency to the other. 
The network shows some obvious connections, like the Euro (EUR) having an impact on the Danish Krone (DKK) exchange rate, and whole contagion paths, like the one starting from the Euro (EUR) to the British Pound (GBP), to the Canadian Dollar (CAD) and ending with an effect on the Mexican Peso (MXN). With the causal networks we can establish where the currency of interest is positioned and, from a systemic risk point of view, where contagion could come from.


\subsubsection{Overall Development of Contagion on Forex} \label{forexcontagion} 

\begin{figure}
    \centering
    \begin{tabular}{c}

     \includegraphics[width=5in]{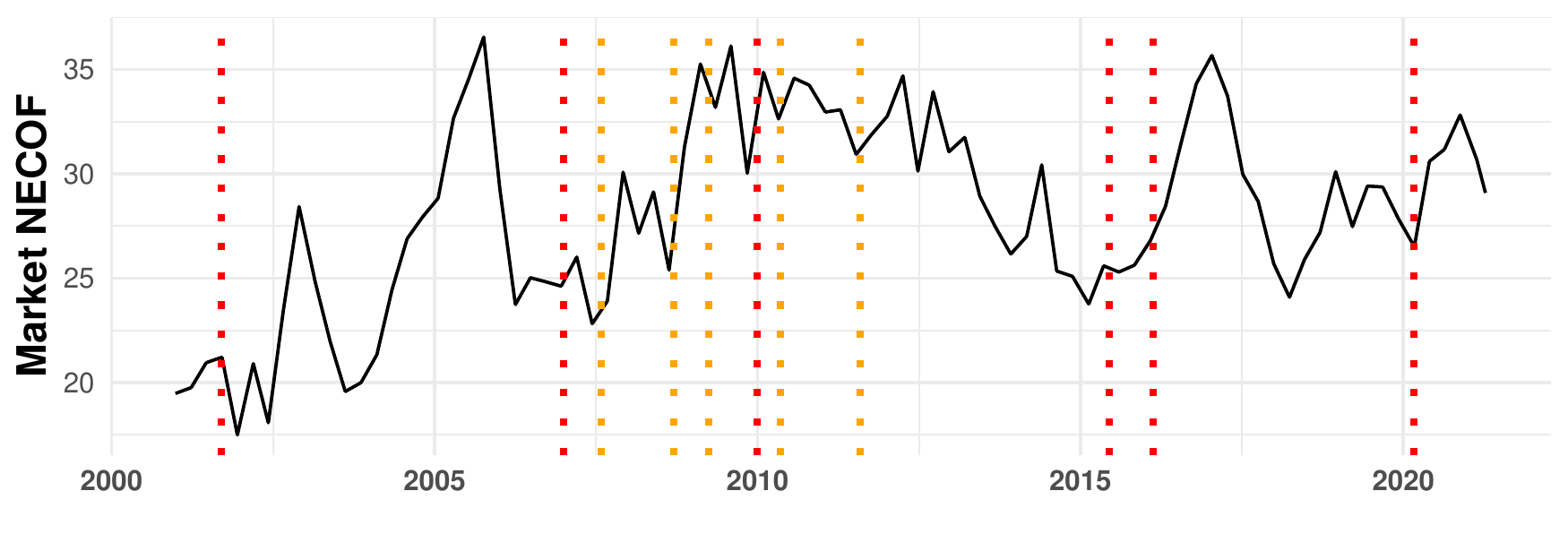}
\\
     \includegraphics[width=5in]{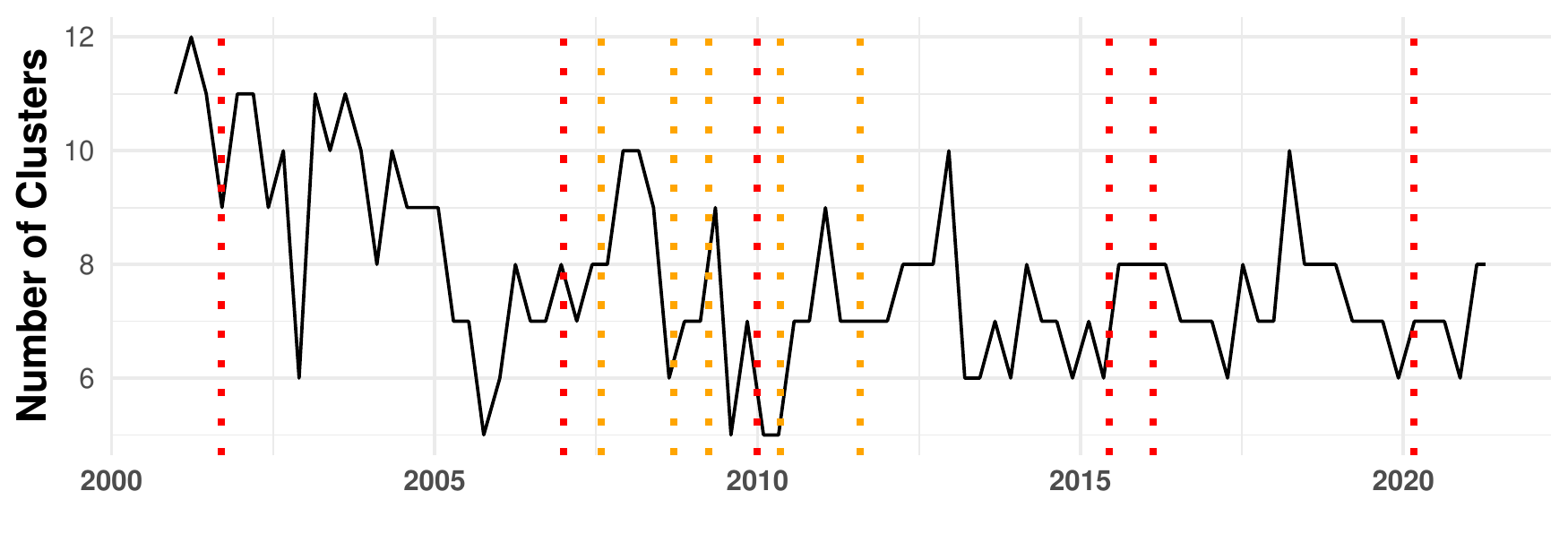}
 \\    
         \includegraphics[width=5in]{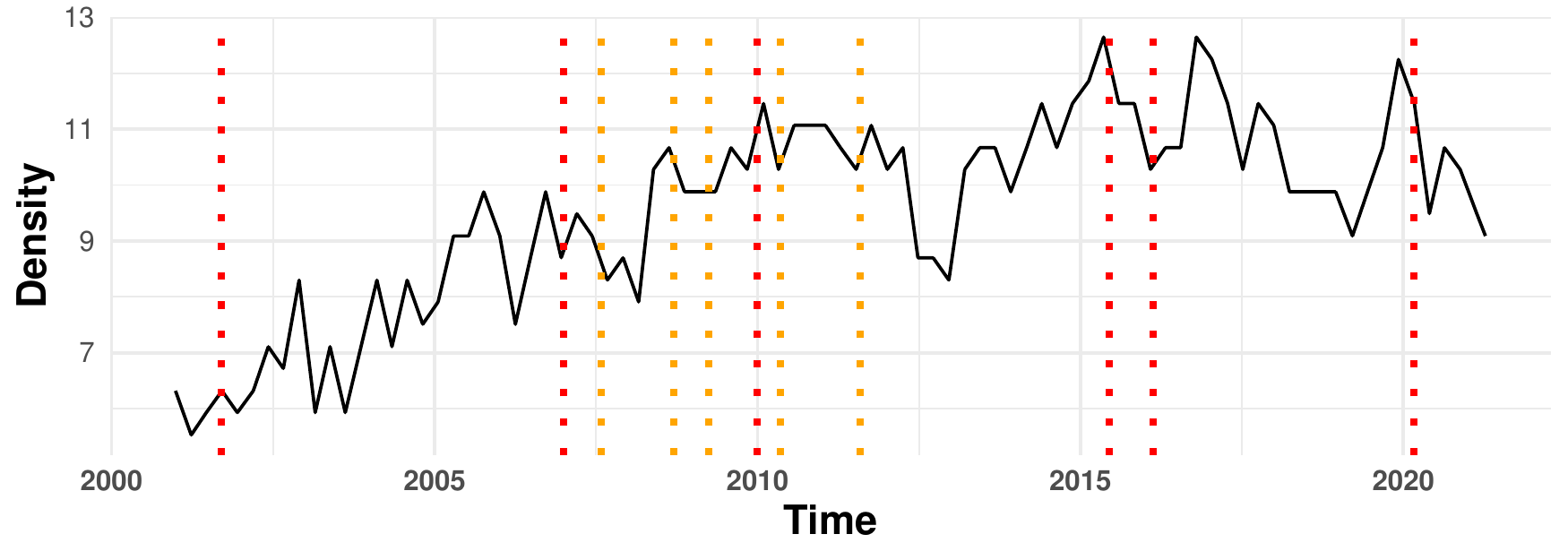}
     
\end{tabular}
    \captionof{figure}[(a) The market NECOF as average over all currencies; (b) The number of Louvain clusters as a measure of clustering; (c) The Density/Connectivity of the the Causal Network.]%
{(a) The market NECOF as average over all currencies; (b) The number of Louvain clusters as a measure of clustering; (c) The Density/Connectivity of the the Causal Network. \par \small We highlight in red the 2001 September 11th Attacks, the 2007-2009 Financial Crisis, the 2010-2012 Sovereign Debt Crisis, the 2015 Chinese stock market crash and the beginning of the Covid Pandemic. In orange we see the most significant events during the timeline of the 2007-2012 global recession period: 9 August 2007 \enquote{The emperor has no clothes!}  BNP Paribas declares publicly that they cannot reasonably value three of their funds due to exposure to the US subprime mortgage market and subsequently freeze the funds; 15 September 2008 investment bank Lehman Brothers is allowed to go bankrupt; 9 May 2010 IMF approves a loan for Greece in a joined rescue package with the EU; 06 August 2011 Standard \& Poor's downgrade America's credit rating from AAA to AA+}
    \label{fig:threegraphs}
  \end{figure}

We estimate our model with a time window of one year (250 business days), rolled over every three months. This allows us to analyse the development of contagion on the Forex from a macro-economic point of view. For each contagion map, we estimate the NECOF at each time period. Figure \ref{fig:threegraphs}(a) shows the evolution of the mean NECOF over time. The contagion rises sharply at the beginning of the 2000s, and then oscillates between 20\% and 35\%. The rise at the beginning is caused by some of the currencies that were formerly pegged to the USD becoming more free and hence more connected to the other currencies on the Forex. This initial rise is interrupted by a peak around sovereign debt crises of low and middle-income countries that began in 2002 as defined by \cite{laeven2018systemic}.

Figure \ref{fig:threegraphs}(b) shows how clustering evolves through time and in response to economical events. The clustering effect is represented by counting the number of clusters present on the network using the Louvain algorithm: the lower the number of clusters, the higher the clustering of the network. The high number of clusters at the beginning of the considered period is driven by the many currencies that were pegged to the USD during the 2000s, whose exchange rates over the USD remained nearly constant. If a currency does not show any variation it will by definition be counted as its own cluster, hence increasing the overall number of clusters detected. As the currencies became more freely traded, in a similar way as the market NECOF rose, so did the number of clusters drop. This clustering effect during a crisis has been reported in many studies on the Forex topology \citep{wang2012similarity,keskin2011topology,kwapien2009analysis,wang2014dynamics}.

Figure \ref{fig:threegraphs}(c) shows that the Forex is predominantly a sparse graph with a relatively low density overall. The effect of globalisation and increased interconnectedness is reflected in the positive trend in the density development over first decade of the 2000s. Once the Financial Crisis began in 2007, however, the density decreased slightly and then levelled off.

All three metric show clear reaction to major economic events, in particular the events of the September 11th 2001 World Trade Center Attacks, the 2007-2009 Financial Crisis, the 2010-2012 European Sovereign Debt Crisis, the 2015 Chinese stock market crash and the beginning of the Covid Pandemic based recession in March of 2020. The charts in Figure \ref{fig:threegraphs} suggest a tendency for the Forex market participants to cluster during a crisis (and so the number of clusters goes down) and for the market contagion to increase. 

\subsubsection{Contagion Clustering in Forex} \label{clusters}

From the previous section we know that there seems to be a tendency for the clustering to increase during a period of economic upheaval. In this section we will examine this dynamic in more detail. In Figure \ref{fig:clusterplots} we show the evolution of the clustering through the 21 years of data. The structure of these clusterplots is similar to that of an adjacency matrix and shows the rows and columns labelled by the different currencies. Instead of showing links or link weights, however the matrix shows how often pairs of currencies belong to the same cluster. The more often two currencies are assigned to the same community by the Louvain algorithm during the specified time frame, the darker and larger the dot connecting the pair gets. If the dot connecting two currencies is a solid dark blue, it means that the pair of currencies were within the same cluster 100\% of the time within the considered period. The only pair that shows such perfect 100\% connection is the Euro (EUR) with the Danish Krone (DKK). Empty cells indicate that the corresponding currencies were never placed in the same cluster within the specified time period

The classically assumed geographically based clusters can be roughly identified within these clusterplots --- but the structure is not so obvious and, more importantly, it changes in reaction to economic events. The main clusters can be roughly divided into a European cluster, an Commonwealth cluster, an Emerging Economies cluster and finally an Asian cluster. Within the European clusters we have some clear oddities. The British Pound  (GBP) switches between the European cluster and the Commonwealth cluster. The Japanese Yen (JPY) is often more connected to the European cluster and specifically the Swiss Franc (CHF), especially so during a crisis. The behaviour of CHF and JPY during a financial crisis is very interesting and will be described in more depth in Section \ref{individualresults}.


As mentioned in Section \ref{forexcontagion} most studies find higher clustering during a crisis. However, from Figure \ref{fig:threegraphs} we know that the number of clusters did not go down immediately during the Financial Crisis of 2007-2009, and that this clustering effect was much more significant during the Sovereign Debt Crisis that followed in 2010. This effect is clearly visible when we compare Figure \ref{fig:clusterplots}(b) and Figure \ref{fig:clusterplots}(c) - the clusters in Figure \ref{fig:clusterplots}(c) (the sovereign debt crisis) are much more defined than the ones in Figure \ref{fig:clusterplots}(b) (the financial crisis of 2007). The European cluster fully de-constructs during this period and does not really ever recover its original structure, unlike the other clusters. The European sovereign debt crisis seems to have caused structural changes to the contagion structure between the Euro and the rest of the European currencies, with effects still lasting to date. These changes cannot really be attributed to the growth and change within the Euro zone as such, because most of these changes took place well before the sovereign debt crisis hit.
The Commonwealth cluster of New Zealand, Australia, Canada, South Africa and sometimes Britain merges during the Financial Crisis with the New Economies cluster of Mexico and Brazil. The clustering effect of the 2007-2010 period can be seen much more clearly when considering changes in clustering during specific time periods than when considering clusters using the full 21-year time series.
 
During the Chinese Market Crash, we find again the typical reaction of higher contagion. The global number of clusters does not change much during this crisis, but Figure \ref{fig:clusterplots}(e) shows some signs of clustering. The European, the Commonwealth and the Emerging Economies clusters are notably more defined. There are some changes within these communities: the Norwegian Krone (NEK) and Swedish Krona (SEK) join the Commonwealth cluster and the Indian Rupee (INR) and the South African Rand (ZAR) move from their respective clusters to the Emerging Economies cluster. It is notable that India, although a member of the Commonwealth, finds itself rarely if at all within the associated cluster.

Unsurprisingly, the Chinese Market Crash seems to have had the largest impact on the structure of the Asian cluster. Even though Figure \ref{fig:clusterplots}(e) covers a relatively short period, the Asian currencies are spread all over the map and almost do not look like a clear cluster at all. During the crisis several of the Asian currencies find more connection with currencies outside of the Asian cluster. The Chinese Yuan Renminbi (CNY), for example, interacts much more with other currencies than in the previous periods. Also, interesting is the behaviour of the Swiss Franc (CHF) during this crisis --- it completely leaves the Euro-centred community and has only connections with Asian currencies, probably due to its traditional role of a safe-haven currency \citep{henderson2006currency, de2015behavior, jaggi2019macroeconomic}.

The period from March 2016 until February 2020 is a period of relative calm, and the clustering resembles the first plot of similar calm, apart of course from the European cluster. One other interesting mention is the Japanese Yen (JPY) that finally moves away from the Asian cluster almost completely.

\begingroup
\setlength{\tabcolsep}{3 pt} 
\renewcommand{\arraystretch}{0.3}

  \begin{minipage}[b]{1\linewidth}
    \centering
    \begin{tabular}{ccc}
     \includegraphics[width=2in]{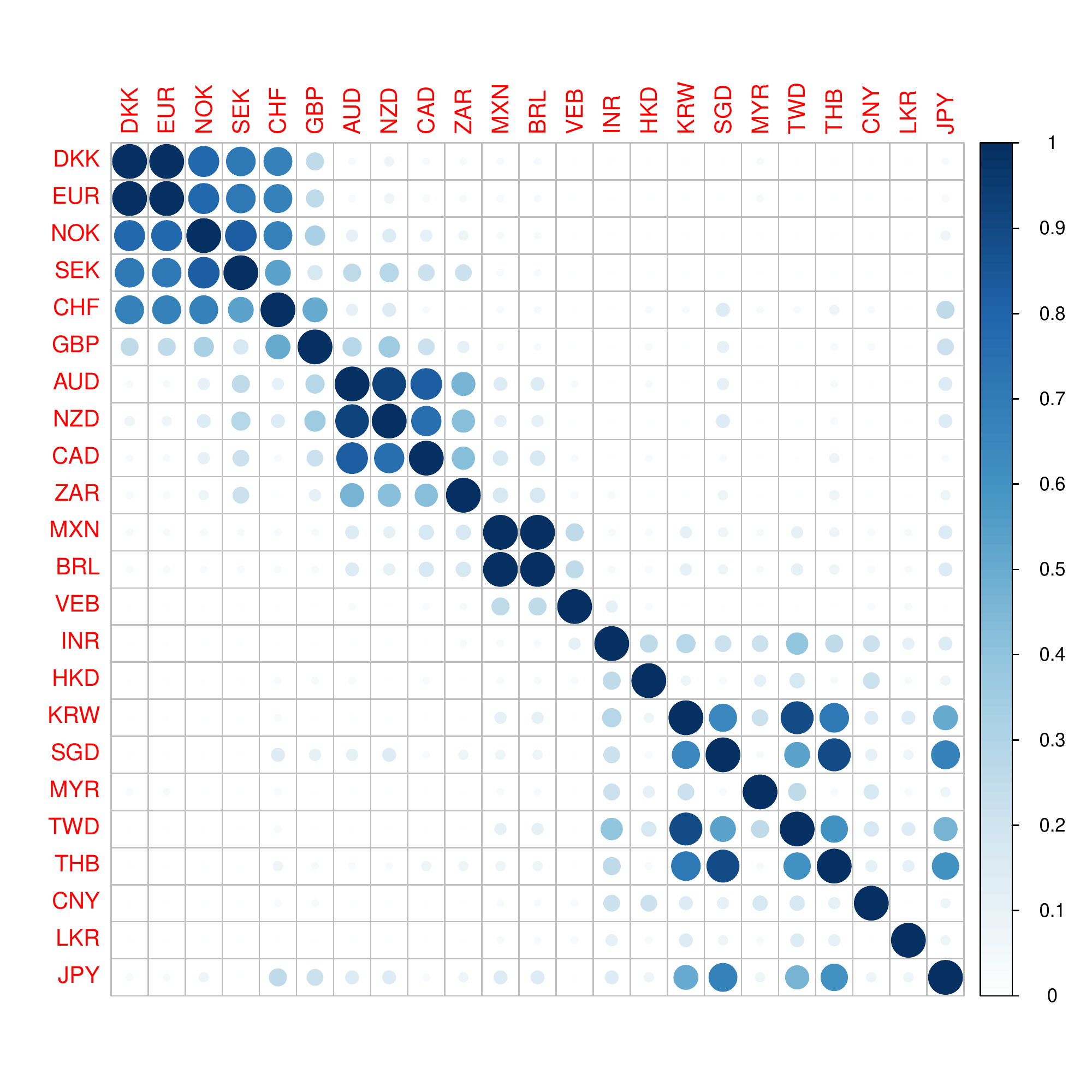}
&
     \includegraphics[width=2in]{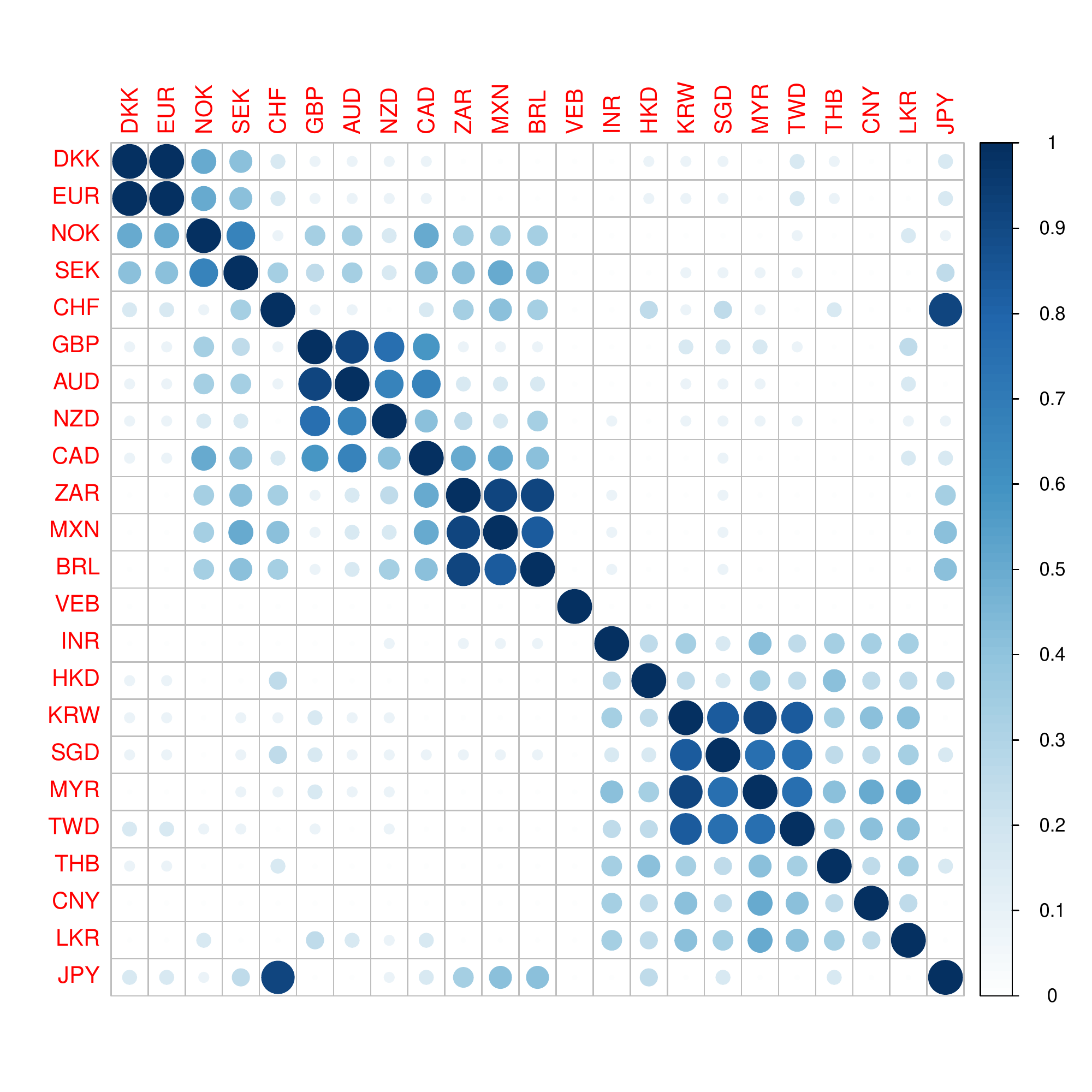}
&
    \includegraphics[width=2in]{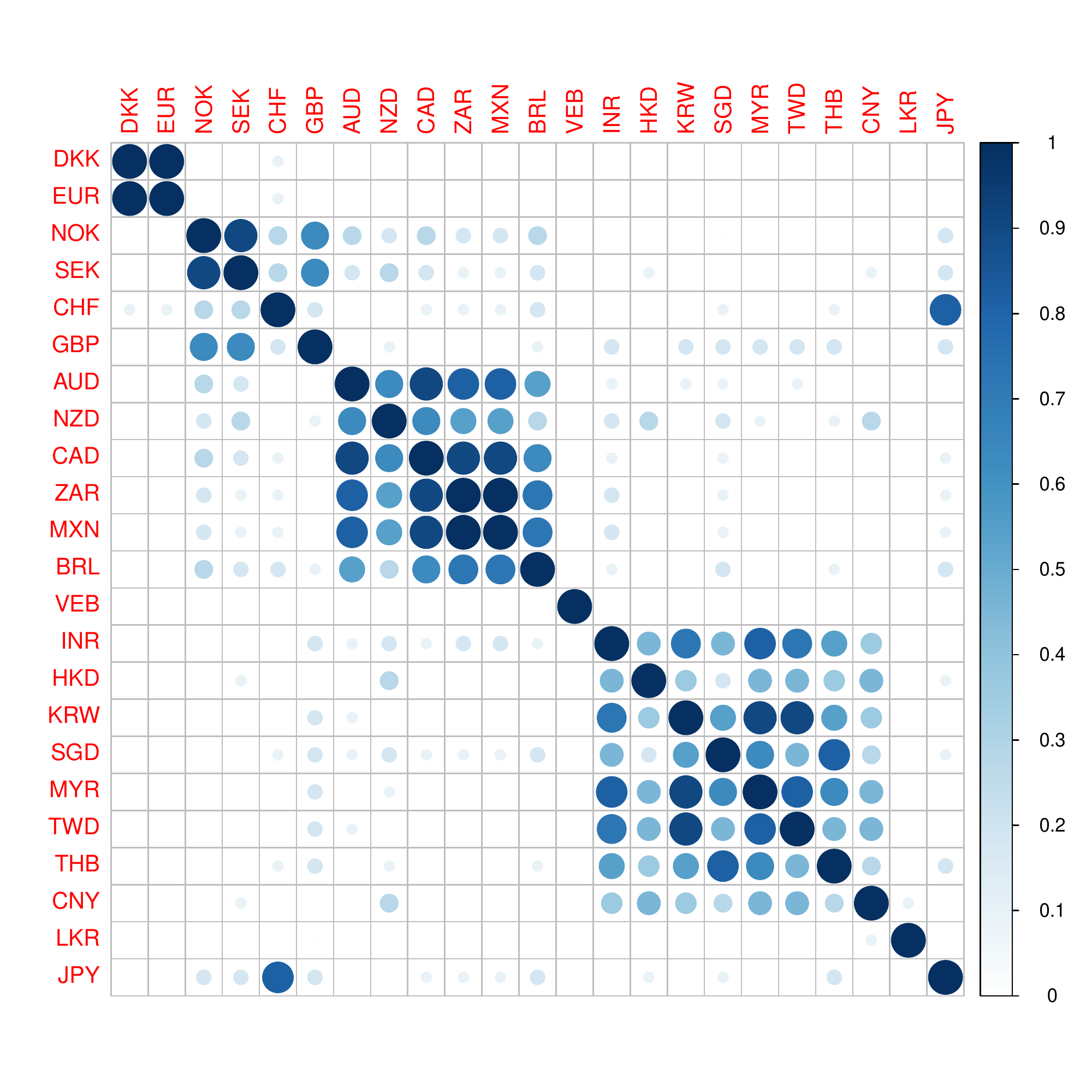}    
\\
	(a) Jan 2000 - Jul 2007
&
	(b) Aug 2007 - Apr 2010
&
(c) May 2010 - Dec 2012
 \\
     \includegraphics[width=2in]{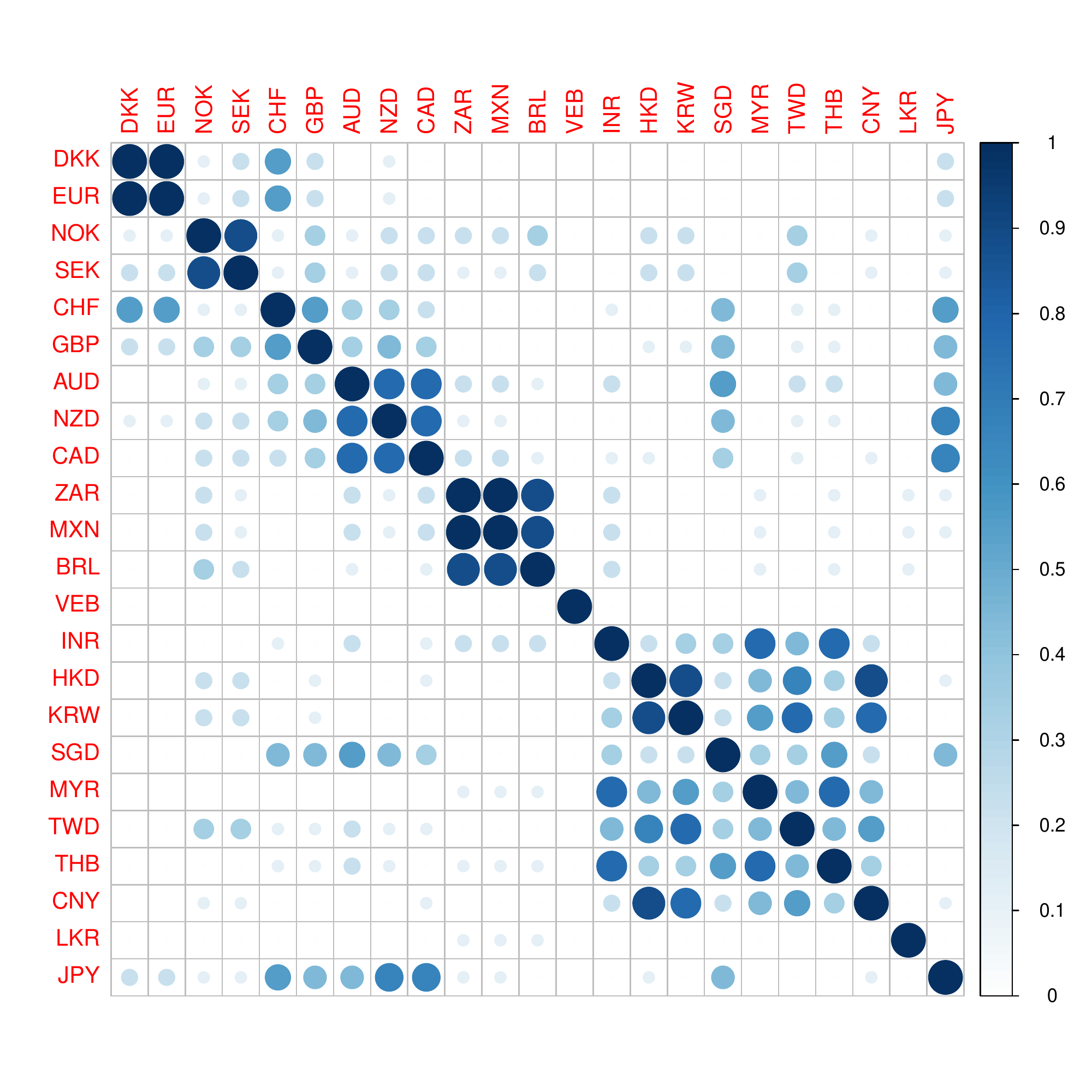} 
&
     \includegraphics[width=2in]{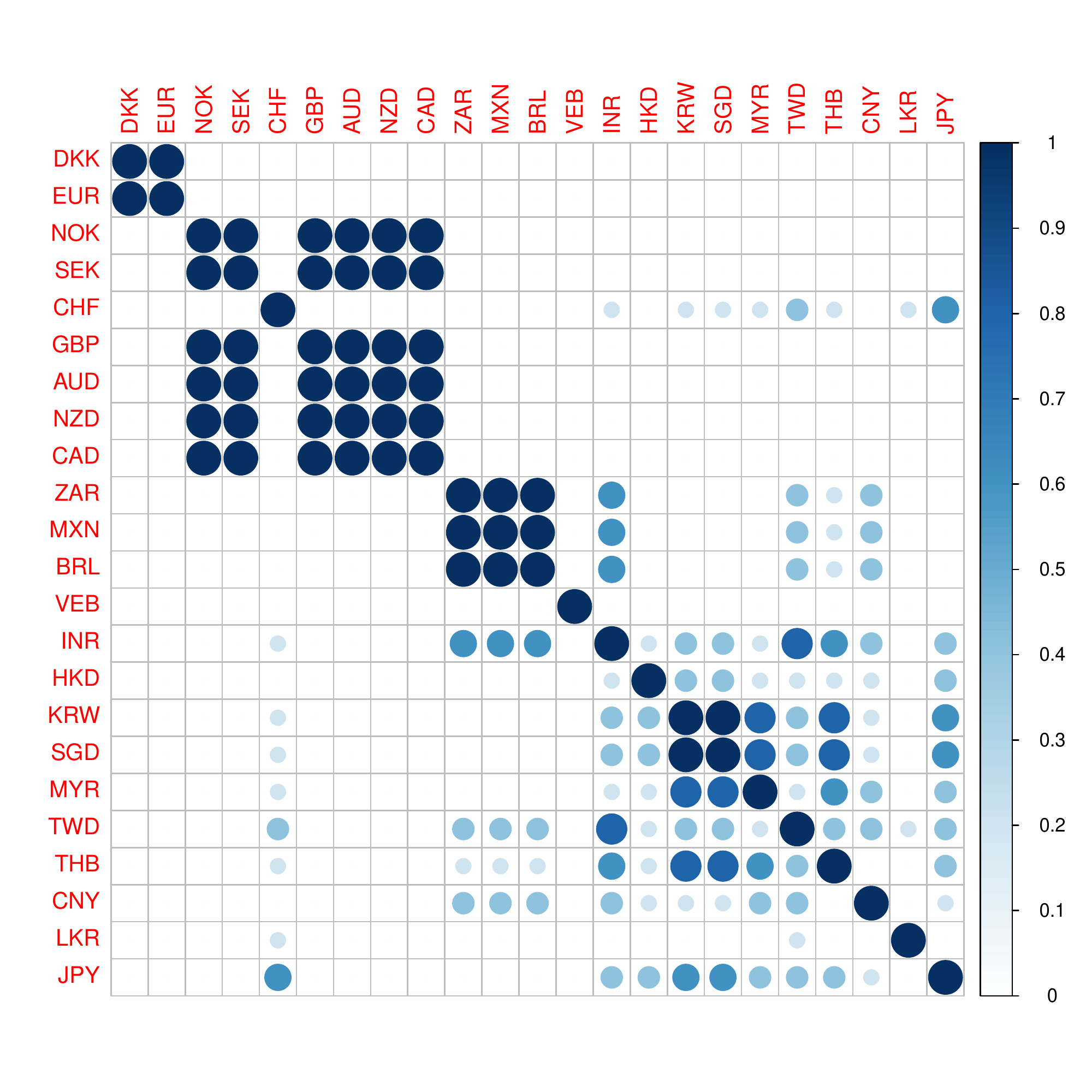}
&
     \includegraphics[width=2in]{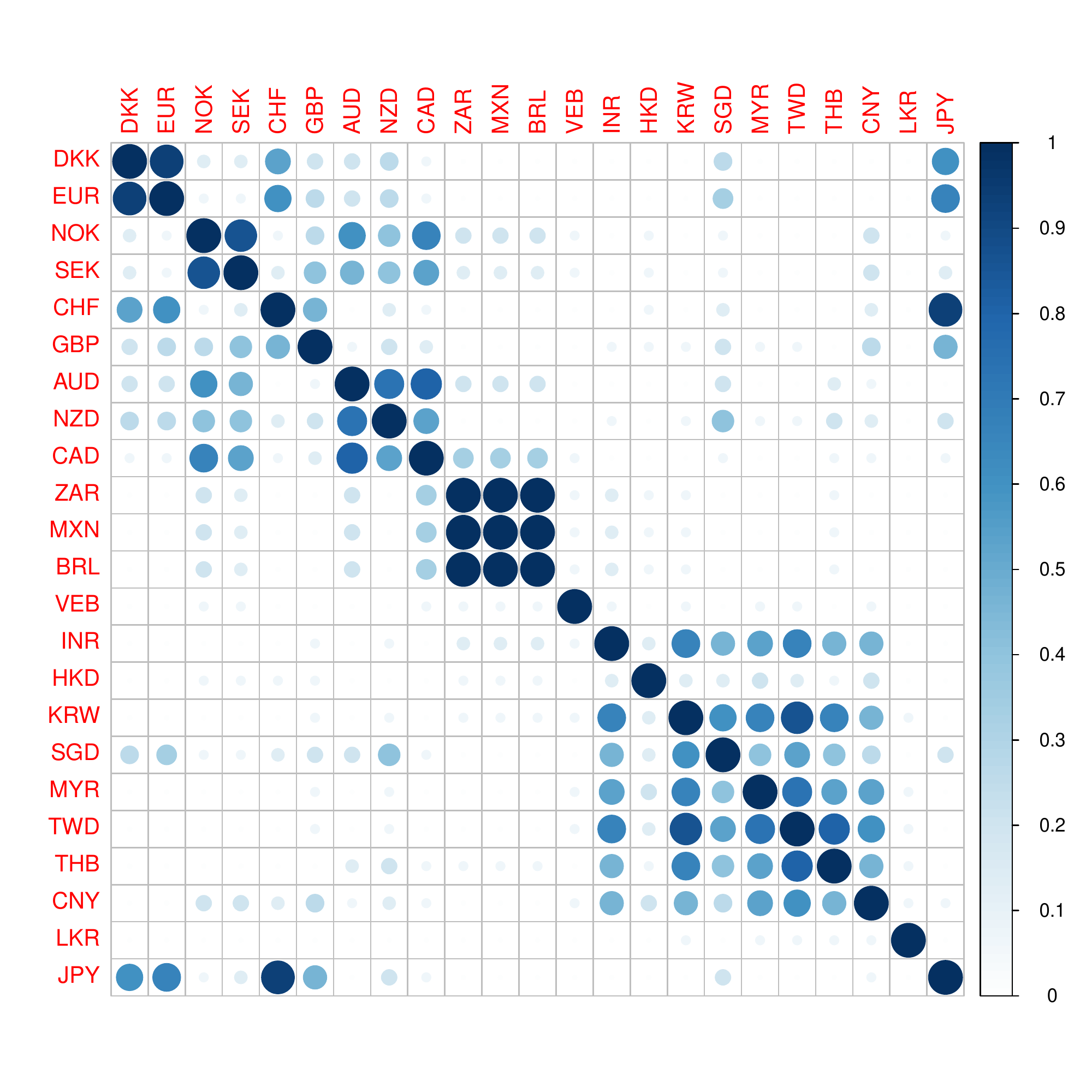}   
\\
	(d) Jan 2013 - May 2015
&
	(e) Jun 2015 - Feb 2016
&
	(f) Mar 2016 - Feb 2020
     
\end{tabular}
    \captionof{figure}[Evolution of the clustering over time and the impact of financial crises.]{These figures show the evolution of the clustering over time and the impact of financial crises. (a) Beginning Period: January 2000 - July 2007; (b) Financial Crisis: August 2007 - April 2010; (c) Sovereign Debt Crisis: May 2010 - December 2012; (d) Intermediate Period January 2013 - May 2015; (e) Chinese Stock Market Crash: June 2015 - February 2016; (f) Post-Crisis Period: March 2016 - February 2020.}
    \label{fig:clusterplots}
  \end{minipage}
\endgroup  
  
In March 2020 the Covid Pandemic becomes a global crisis, and the Covid Recession officially starts. We see from Figure \ref{fig:threegraphs} that the market contagion increased immediately at the beginning of the pandemic, but the number of clusters did not change very much. What changed, as in the previous crisis, is the redistribution of the currencies within the clusters themselves --- but in a different and more dramatic fashion. Figure \ref{fig:covidcluster} shows this redistribution. The network is much more compact and in fact the density does sharply go down, i.e. the network has fewer links. This means we have new clustering with a lower number of links, but more significant links that lead to a much higher contagion on the markets. The  CNY and the INR join the European cluster, whereas the GBP, SEK and NOK move together to join the Commonwealth cluster. That the impact of the Covid Pandemic on the contagion map and clustering within the Forex would be somewhat different from the previous financial crisis was to be expected --- even the markets reacted very differently. The stock markets fell faster than ever before\footnote{For example the S\&P 500 index fell by 34\% between Feb. 19 and March 23, which constitutes the fastest fall in market in history, for further details see \cite{roubini2020coronavirus}.} and the impact of the Covid-19 recession seems to be having an impact on the global structure of the world economy \citep{carlsson2020coronavirus, sulkowski2020covid}. The cross-border financial interventions and foreign aid spending reached unprecedented levels during the Covid Pandemic \citep{OECD20210413}. All of these could also explain changes within the clustering.

  \begin{minipage}[b]{1\linewidth}
    \centering
    \begin{tabular}{cc}
\includegraphics[width=2in]{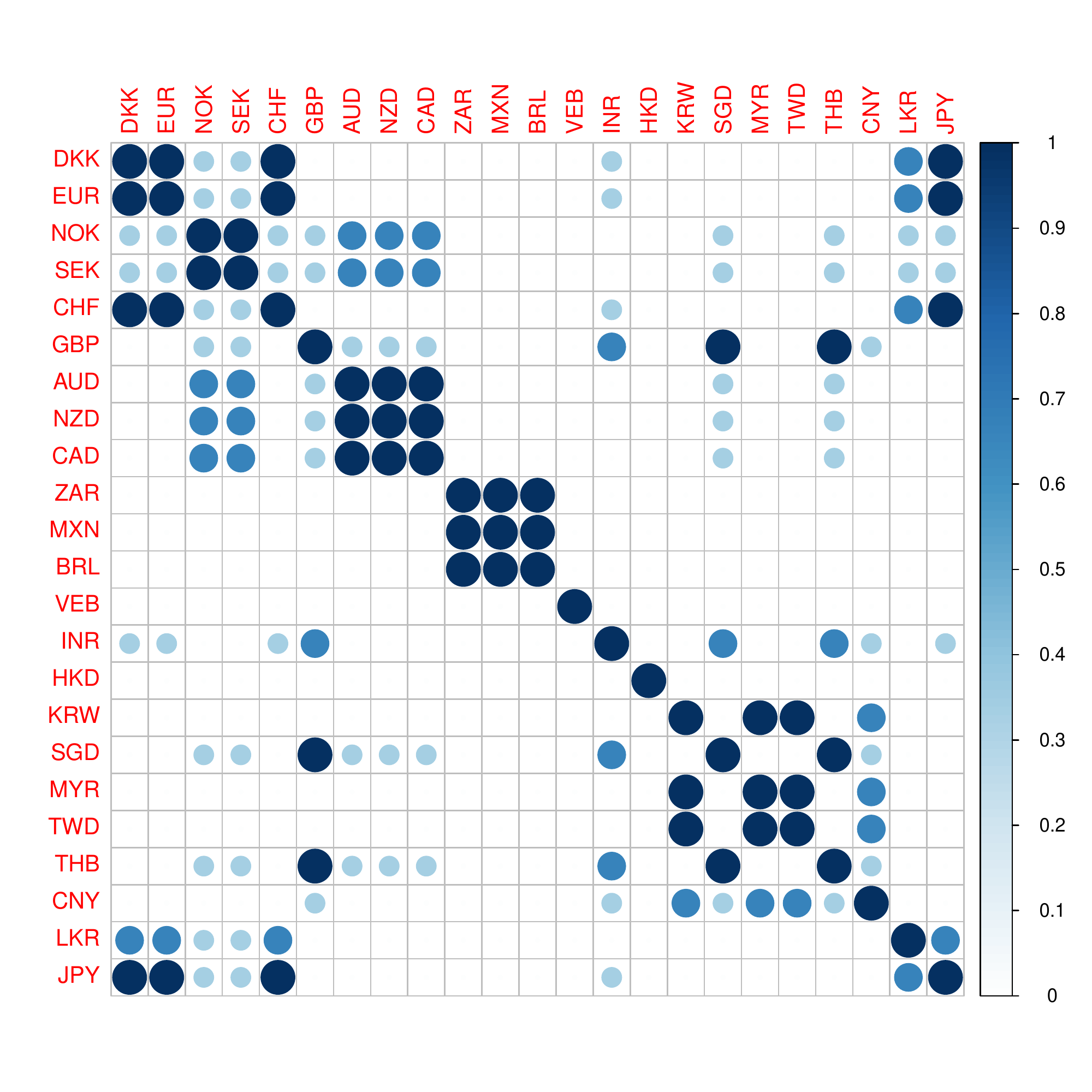}   
&
\includegraphics[width=0.6\textwidth]{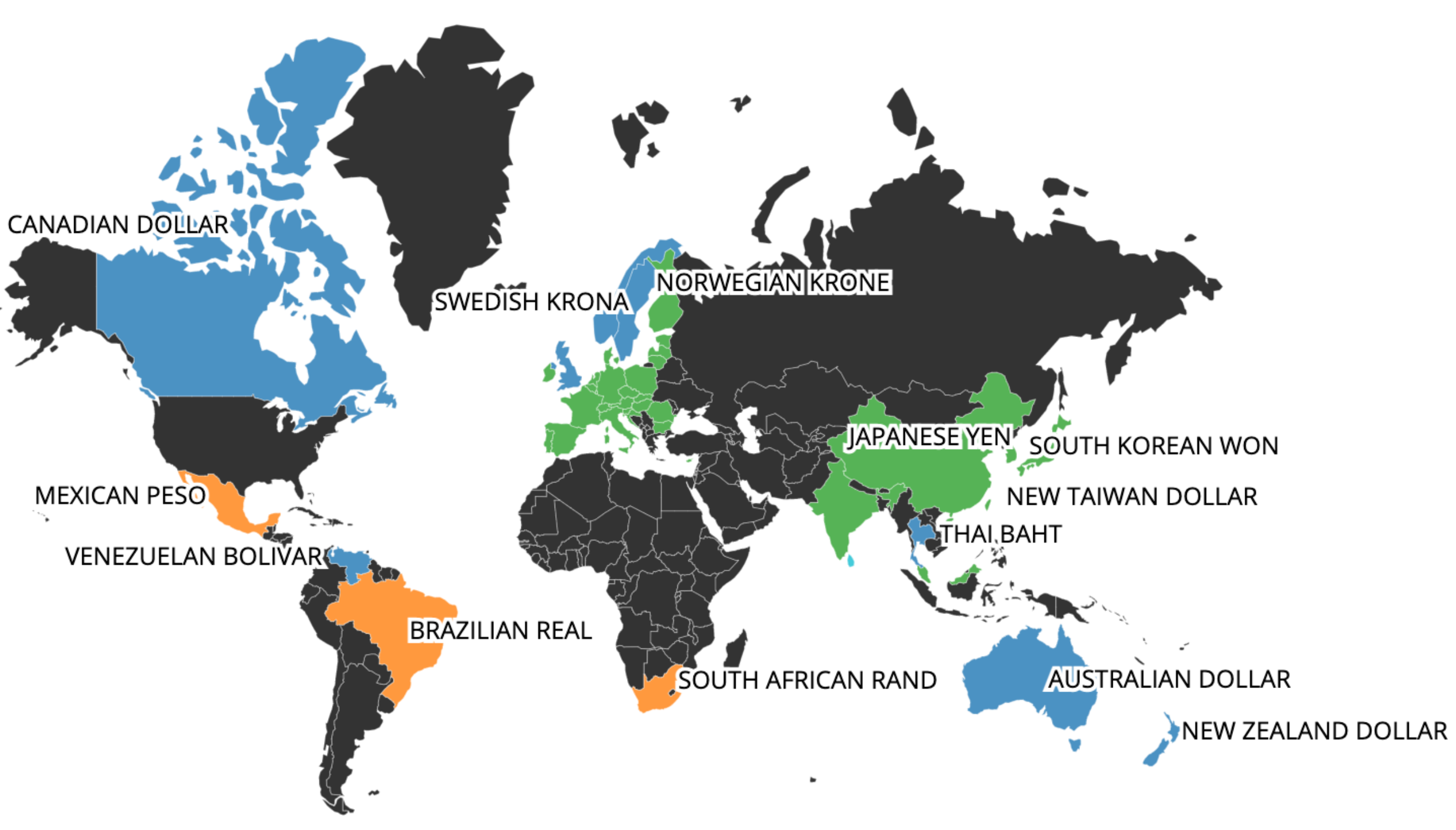}   
\\
	Mar 2020 - Apr 2021
&
\end{tabular}
\captionof{figure}[Impact of the Covid Recession on the contagion-based clustering, for the period starting in March 2020.]{Impact of the Covid Recession on the contagion-based clustering, for the period starting in March 2020. The countries in black are not part of the analysis.}
\label{fig:covidcluster}
\end{minipage}



  \begin{minipage}[b]{1\linewidth}
    \centering
    \begin{tabular}{c}
 
     \includegraphics[width=4.5in]{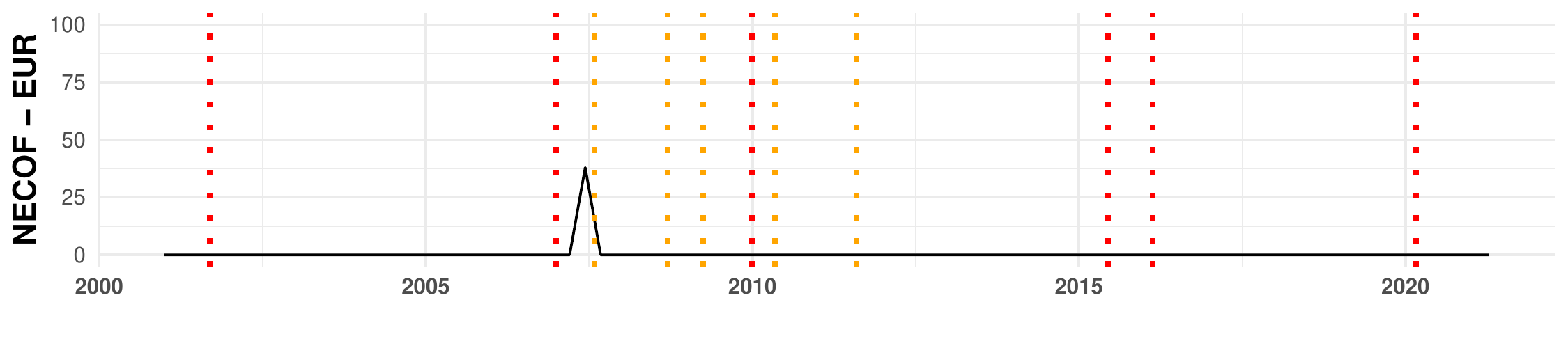}
\\
     \includegraphics[width=4.5in]{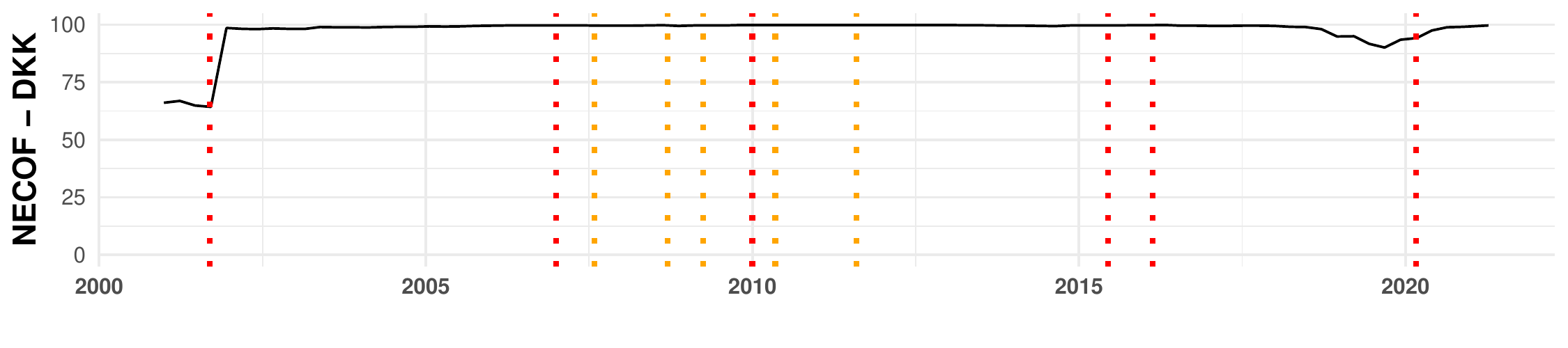}
\\    
      \includegraphics[width=4.5in]{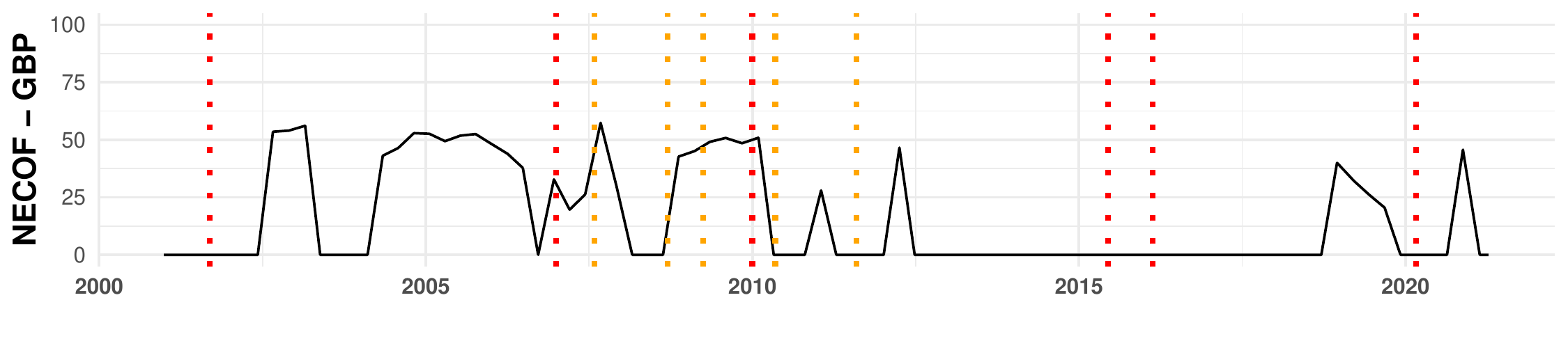}
 \\    
       \includegraphics[width=4.5in]{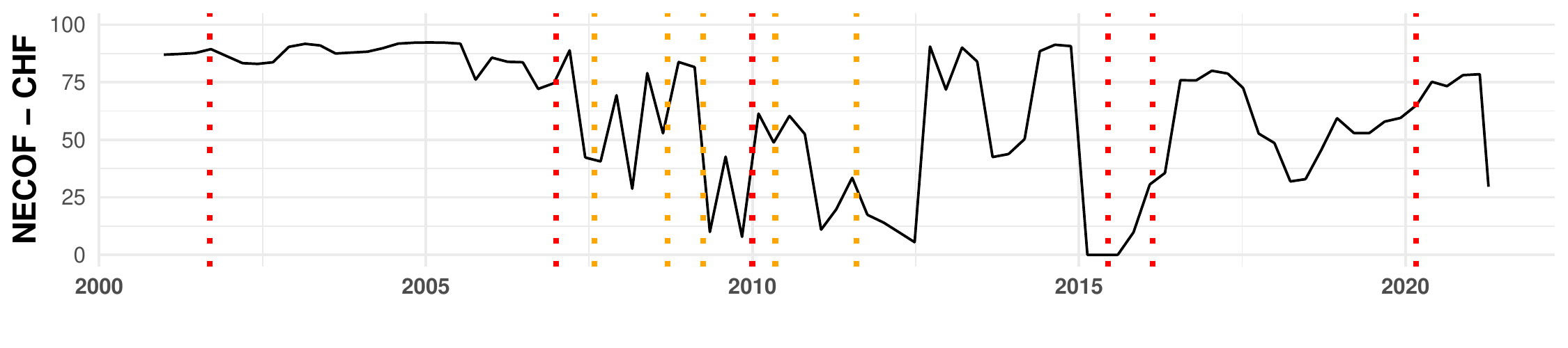}   
       \\    
       \includegraphics[width=4.5in]{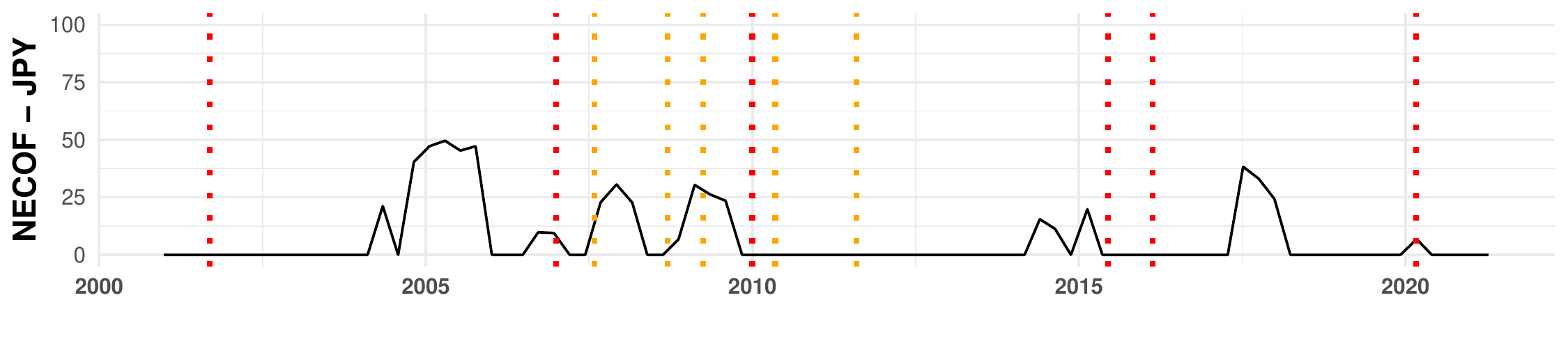}   
 \\    
       \includegraphics[width=4.5in]{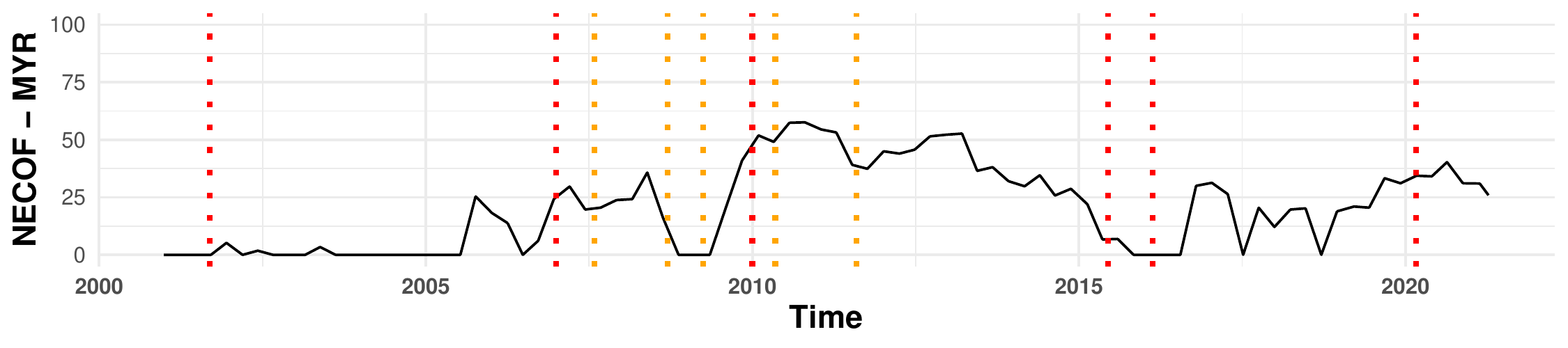}          
\end{tabular}
\captionof{figure}[NECOFs for individual currencies.]%
{NECOFs for individual currencies. \par \small Red and orange vertical lines represent the same events as in Figure \ref{fig:threegraphs}.}
    \label{fig:individualNECOFs}
  \end{minipage}

\subsubsection{Individual Network Contagion Dynamics} \label{individualresults}
In this section we will look in more detail at some interesting behaviour of individual currencies. In Figure \ref{fig:individualNECOFs} we show how the NECOF contagion factors evolve for the Euro (EUR), the Danish Krone (DKK), the British Punds (GBP), the Swiss Franc (CHF) and the Malaysian Ringgit (MYR).

\begin{figure}[tb] 
\centering
\includegraphics
[width=0.6\textwidth]{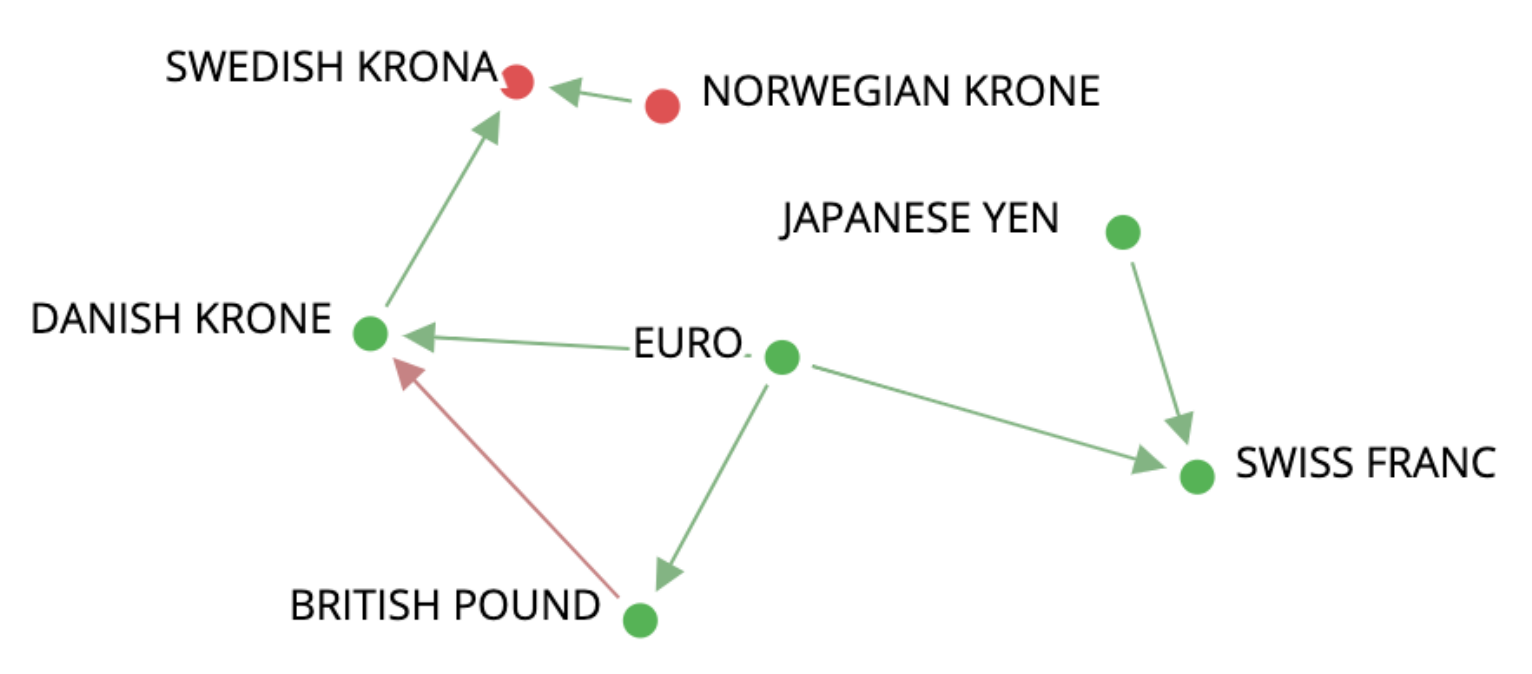} 
\caption{Subgraph of the Causal Network for the GBP in 2019.}
\label{fig:2019GBP}
\end{figure}

We previously discussed the pair EUR and DKK and their very high correlation. Considering the causality links, we additionally see that it is the EUR that influences the DKK and not vice versa, as shown in Figure \ref{fig:2019GBP} and \ref{fig:heatmap}(b). This unsurprising fact shows that the NECOF measure is much more meaningful than a correlation analysis in describing the risk of contagion of a financial asset. Two currencies can have a very high correlation and yet completely opposite NECOFs. In fact, the DKK shows a NECOF of almost 100\% most of the time, while the NECOF of the EUR is usually at 0\% --- increasing significantly only once at the beginning of the Financial Crisis in 2007. This causal relationship between the EUR and the DKK holds throughout the considered 21 years.
%

The next interesting example of a NECOF path is that of the GBP. In the previous section we saw that the GBP has a tendency to switch clusters between Europe and the Commonwealth cluster. It usually is a currency that influences others and generally has a low NECOF around 0\% --- this is in line with what \cite{giudici2018corisk} find for the United Kingdom based on Corporate Default Swap spreads (CDS). In our study however, we find that the GBP's NECOF has a tendency to rise sharply during turbulent times like the Financial Crisis. Curiously, the GBP was hardly ever under the influence, from a contagion point of view, of the EUR ---  unless, again, there was a crisis. The GBP experienced contagion from the EUR in 2007, during the financial crisis, and then during Brexit, around the time when the first draft withdrawal agreement was negotiated and endorsed by the EU members at the end of 2019, as seen in Figure \ref{fig:2019GBP}.

The CHF is traditionally considered a safe-haven currency \citep{henderson2006currency, de2015behavior, jaggi2019macroeconomic}. The CHF presents a relatively volatile and often high NECOF. This is in contrast to the assumed safety of the CHF, but on the other hand the NECOF of the CHF has a tendency to go down precisely in times of crisis, which again speaks for the safe haven status. The low NECOFs during the European Sovereign Debt Crisis are an example of this behaviour. The sharp fall of the NECOF at the end of 2014 is due to the intervention of the Swiss central bank, which tried to peg the currency to the EUR to prevent the increase in value of CHF. This attempt was however scrapped in January of 2015. And after the Chinese market crash passed, the NECOF shot up again.

In the past, the JPY was seen more \enquote{as a low interest rate or funding currency}\citep{henderson2006currency}, but most recent studies classify it a safe-haven currency \citep{botman2013yensafe, jaggi2019macroeconomic}. As with the CHF, the JPY seems to appreciate in value during a crisis and during high volatility periods \citep{ranaldo2010safe}. As Figure \ref{fig:individualNECOFs}  shows, the JPY presents a very low NECOF for most of the time, and this would seem to validate the consideration of the JPY as a safe-haven currency. We further find a dependency of the CHF on the JPY, illustrated in Figure \ref{fig:2019GBP}. The arrow goes from the JPY to the CHF in 75\% of the networks (never in the opposite direction), which indicates that there is contagion that goes form the JPY to the CHF. The causal effect from the JPY to the CHF and the lower NECOF of the JPY in general would suggest that the JPY could be even considered  a better safe haven than the CHF. This is exactly what \cite{fatum2016intra} and \cite{de2015behavior} find when comparing these two currencies in terms of safe-haven characteristics.

The last currency we will consider in detail is the MYR. The MYR shows the NECOF evolution of an Asian currency that was pegged to the USD until the 2005. The NECOF falls in 2008 and after the 2015 Chines stock market crash, mainly because of the intervention by the Bank Negara (the Central Bank of Malaysia) to prevent the exchange rate over the USD from plummeting. At times of important intervention by a central bank the contagion from other currencies clearly decreases. The Malaysian economy and financial sector have grown  during the 20 years considered here, and it is interesting how resilient to contagion it seems to be, unlike the other south Asian currencies.

\section{Conclusion}
\label{conclusion}

Financial contagion measures have always had causal aspirations. In this work we have unified this concept of causality with a defined measure that allows for a quantifiable causal interpretation of contagion relationships on the Forex market. We corroborate and extend results from different studies within one unifying framework and are able to answer very practical questions like how contagion spreads on the Forex Market and which currencies are at the highest risk of contagion at any given time. 

We have recreated a series of causal networks based on 23 exchange rates over the USD, spanning over 21 years, and present both the overall development of contagion on the Forex as well as individual network contagion dynamics. We have shown how to read these causal networks as contagion maps to pinpoint sources of contagion and how the contagion paths on the Forex evolve through time. We were able to identify a new promising group of financial indicators that take contagion and systemic risk directly into account. The newly identified measure of network contagion (NECO) seems to be of value for both a market level evaluation and the analysis of single financial instruments alike. We discussed the network contagion factors' (NECOF) evolution for a subset of currencies, to demonstrate how this metric can be used to identify and evaluate a financial instrument from an investment and hedging point of view.

In contrast to the correlation networks, we obtain causal directions. Because they are inherently spares, causal networks do not have to be filtered, e.g. via Minimum Spanning Tree (MST) methods \citep{mantegna1999hierarchical}, and can be easily analysed and evaluated. \cite{kazemilari2013analysis} compare the use of different centrality measures in constructing MSTs for the Forex, (including degree,
betweenness, closeness and eigenvector). Filtering is found to be important, but results are sensitive to the exact measure of correlation used as well as the distance measure. There are other filtering methods and so the choice of the  method itself will also affect results \citep{cook2016network, tumminello2005tool, serrano2009extracting, marcaccioli2019polya}.

The application of causal graphical models to financial data is in its infancy and there are still interesting challenges. In our case we assumed normality for log-returns and found our sample to be stationary, but a model that could deal automatically with non-normality, fat-tails, heteroskedasticity and non-stationarity of the data would be advantageous for further applications in finance. 
The trading on the Forex is active 24 hours a day \citep{goodhart1993central} and so we were able to use prices for all currencies at the same time instant. This is not the case for most other markets and assets being traded, and hence the impact of asynchronously observed returns on the analysis would have to be taken into account \citep{burns1998correlations}. Whereas the contemporaneous contagion is the most significant in a fast moving and liquid market, it could also be of interest to consider economic cycles and longer time delays. Lastly, any measurable confounder can be added into our model, should we want to see how other variables, e.g. interest rates or inflation, impact the Forex contagion. For unmeasured confounders and latent variables the The Fast Causal Inference algorithm (FCI) \citep{spirtes1995causal,spirtes2000causation} can be used.




%

\bibliographystyle{chicago}
\bibliography{biblio}

\begin{thebibliography}{}

\bibitem[\protect\citeauthoryear{Avdjiev, Giudici, and Spelta}{Avdjiev
  et~al.}{2019}]{avdjiev2019measuring}
Avdjiev, S., P.~Giudici, and A.~Spelta (2019).
\newblock Measuring contagion risk in international banking.
\newblock {\em Journal of Financial Stability\/}~{\em 42}, 36--51.

\bibitem[\protect\citeauthoryear{{Bank of international Settlement}}{{Bank of
  international Settlement}}{2019}]{bank2019triennial}
{Bank of international Settlement} (2019).
\newblock Triennial central bank survey.

\bibitem[\protect\citeauthoryear{Barigozzi and Hallin}{Barigozzi and
  Hallin}{2017}]{barigozzi2017network}
Barigozzi, M. and M.~Hallin (2017).
\newblock A network analysis of the volatility of high dimensional financial
  series.
\newblock {\em Journal of the Royal Statistical Society: Series C (Applied
  Statistics)\/}~{\em 66\/}(3), 581--605.

\bibitem[\protect\citeauthoryear{Barigozzi, Hallin, Soccorsi, and von
  Sachs}{Barigozzi et~al.}{2020}]{barigozzi2020time}
Barigozzi, M., M.~Hallin, S.~Soccorsi, and R.~von Sachs (2020).
\newblock Time-varying general dynamic factor models and the measurement of
  financial connectedness.
\newblock {\em Journal of Econometrics\/}.

\bibitem[\protect\citeauthoryear{Battiston and Martinez-Jaramillo}{Battiston
  and Martinez-Jaramillo}{2018}]{battiston2018financial}
Battiston, S. and S.~Martinez-Jaramillo (2018).
\newblock Financial networks and stress testing: Challenges and new research
  avenues for systemic risk analysis and financial stability implications.
\newblock {\em Journal of Financial Stability\/}~{\em 35}, 6--16.

\bibitem[\protect\citeauthoryear{Blondel, Guillaume, Lambiotte, and
  Lefebvre}{Blondel et~al.}{2008}]{blondel2008fast}
Blondel, V.~D., J.-L. Guillaume, R.~Lambiotte, and E.~Lefebvre (2008).
\newblock Fast unfolding of communities in large networks.
\newblock {\em Journal of statistical mechanics: theory and experiment\/}~{\em
  2008\/}(10), P10008.

\bibitem[\protect\citeauthoryear{Bollen and Pearl}{Bollen and
  Pearl}{2013}]{bollen2013eight}
Bollen, K.~A. and J.~Pearl (2013).
\newblock Eight myths about causality and structural equation models.
\newblock In {\em Handbook of causal analysis for social research}, pp.\
  301--328. Springer.

\bibitem[\protect\citeauthoryear{Botman, de~Carvalho~Filho, and Lam}{Botman
  et~al.}{2013}]{botman2013yensafe}
Botman, M. D.~P., M.~I.~E. de~Carvalho~Filho, and M.~W.~W. Lam (2013).
\newblock {\em The curious case of the yen as a safe haven currency: a forensic
  analysis}.
\newblock International Monetary Fund.

\bibitem[\protect\citeauthoryear{Burns, Engle, and Mezrich}{Burns
  et~al.}{1998}]{burns1998correlations}
Burns, P., R.~Engle, and J.~Mezrich (1998).
\newblock Correlations and volatilities of asynchronous data.
\newblock {\em Journal of Derivatives\/}~{\em 5\/}(4), 7.

\bibitem[\protect\citeauthoryear{Calvo and Reinhart}{Calvo and
  Reinhart}{1996}]{calvo1996capital}
Calvo, S.~G. and C.~M. Reinhart (1996).
\newblock Capital flows to latin america: is there evidence of contagion
  effects?
\newblock {\em World Bank Policy Research Working Paper\/}~(1619).

\bibitem[\protect\citeauthoryear{Carlsson-Szlezak, Reeves, and
  Swartz}{Carlsson-Szlezak et~al.}{2020}]{carlsson2020coronavirus}
Carlsson-Szlezak, P., M.~Reeves, and P.~Swartz (2020).
\newblock What coronavirus could mean for the global economy.
\newblock {\em Harvard Business Review\/}~{\em 3}, 1--10.

\bibitem[\protect\citeauthoryear{Claessens and Forbes}{Claessens and
  Forbes}{2013}]{claessens2013international}
Claessens, S. and K.~Forbes (2013).
\newblock {\em International financial contagion}.
\newblock Springer Science \& Business Media.

\bibitem[\protect\citeauthoryear{Collins and Biekpe}{Collins and
  Biekpe}{2003}]{collins2003contagion}
Collins, D. and N.~Biekpe (2003).
\newblock Contagion: a fear for african equity markets?
\newblock {\em Journal of Economics and Business\/}~{\em 55\/}(3), 285--297.

\bibitem[\protect\citeauthoryear{Colombo and Maathuis}{Colombo and
  Maathuis}{2014}]{colombo2014order}
Colombo, D. and M.~H. Maathuis (2014).
\newblock Order-independent constraint-based causal structure learning.
\newblock {\em J. Mach. Learn. Res.\/}~{\em 15\/}(1), 3741--3782.

\bibitem[\protect\citeauthoryear{Dahlhaus and Eichler}{Dahlhaus and
  Eichler}{2003}]{dahlhaus2003vcausality}
Dahlhaus, R. and M.~Eichler (2003).
\newblock Vcausality and graphical models for time series. v in: P. green, n.
  hjort, and s. richardson (eds.), highly structured stochastic systems.

\bibitem[\protect\citeauthoryear{De~Bock and de~Carvalho~Filho}{De~Bock and
  de~Carvalho~Filho}{2015}]{de2015behavior}
De~Bock, R. and I.~de~Carvalho~Filho (2015).
\newblock The behavior of currencies during risk-off episodes.
\newblock {\em Journal of International Money and Finance\/}~{\em 53},
  218--234.

\bibitem[\protect\citeauthoryear{Diebold and Y{\i}lmaz}{Diebold and
  Y{\i}lmaz}{2014}]{diebold2014network}
Diebold, F.~X. and K.~Y{\i}lmaz (2014).
\newblock On the network topology of variance decompositions: Measuring the
  connectedness of financial firms.
\newblock {\em Journal of Econometrics\/}~{\em 182\/}(1), 119--134.

\bibitem[\protect\citeauthoryear{Edwards}{Edwards}{2000}]{edwards2000contagion}
Edwards, S. (2000).
\newblock Contagion.
\newblock In {\em World Economy}. Citeseer.

\bibitem[\protect\citeauthoryear{Eichler}{Eichler}{2007}]{eichler2007granger}
Eichler, M. (2007).
\newblock Granger causality and path diagrams for multivariate time series.
\newblock {\em Journal of Econometrics\/}~{\em 137\/}(2), 334--353.

\bibitem[\protect\citeauthoryear{Elliott, Golub, and Jackson}{Elliott
  et~al.}{2014}]{elliott2014financial}
Elliott, M., B.~Golub, and M.~O. Jackson (2014).
\newblock Financial networks and contagion.
\newblock {\em American Economic Review\/}~{\em 104\/}(10), 3115--53.

\bibitem[\protect\citeauthoryear{Fatum and Yamamoto}{Fatum and
  Yamamoto}{2016}]{fatum2016intra}
Fatum, R. and Y.~Yamamoto (2016).
\newblock Intra-safe haven currency behavior during the global financial
  crisis.
\newblock {\em Journal of International Money and Finance\/}~{\em 66}, 49--64.

\bibitem[\protect\citeauthoryear{Fenn, Porter, Mucha, McDonald, Williams,
  Johnson, and Jones}{Fenn et~al.}{2012}]{fenn2012dynamical}
Fenn, D.~J., M.~A. Porter, P.~J. Mucha, M.~McDonald, S.~Williams, N.~F.
  Johnson, and N.~S. Jones (2012).
\newblock Dynamical clustering of exchange rates.
\newblock {\em Quantitative Finance\/}~{\em 12\/}(10), 1493--1520.

\bibitem[\protect\citeauthoryear{Forbes and Rigobon}{Forbes and
  Rigobon}{2002}]{forbes2002no}
Forbes, K.~J. and R.~Rigobon (2002).
\newblock No contagion, only interdependence: measuring stock market
  comovements.
\newblock {\em The journal of Finance\/}~{\em 57\/}(5), 2223--2261.

\bibitem[\protect\citeauthoryear{Gai, Haldane, and Kapadia}{Gai
  et~al.}{2011}]{gai2011complexity}
Gai, P., A.~Haldane, and S.~Kapadia (2011).
\newblock Complexity, concentration and contagion.
\newblock {\em Journal of Monetary Economics\/}~{\em 58\/}(5), 453--470.

\bibitem[\protect\citeauthoryear{Giudici and Parisi}{Giudici and
  Parisi}{2018}]{giudici2018corisk}
Giudici, P. and L.~Parisi (2018).
\newblock Corisk: Credit risk contagion with correlation network models.
\newblock {\em Risks\/}~{\em 6\/}(3), 95.

\bibitem[\protect\citeauthoryear{Glasserman and Young}{Glasserman and
  Young}{2015}]{glasserman2015likely}
Glasserman, P. and H.~P. Young (2015).
\newblock How likely is contagion in financial networks?
\newblock {\em Journal of Banking \& Finance\/}~{\em 50}, 383--399.

\bibitem[\protect\citeauthoryear{Goodhart and Hesse}{Goodhart and
  Hesse}{1993}]{goodhart1993central}
Goodhart, C.~A. and T.~Hesse (1993).
\newblock Central bank forex internvention assessed in continous time.
\newblock {\em Journal of international money and finance\/}~{\em 12\/}(4),
  368--389.

\bibitem[\protect\citeauthoryear{Hansen}{Hansen}{2021}]{hansen2021financial}
Hansen, K.~B. (2021).
\newblock Financial contagion: problems of proximity and connectivity in
  financial markets.
\newblock {\em Journal of Cultural Economy\/}, 1--15.

\bibitem[\protect\citeauthoryear{Henderson}{Henderson}{2006}]{henderson2006currency}
Henderson, C. (2006).
\newblock {\em Currency strategy: The practitioner's guide to currency
  investing, hedging and forecasting}.
\newblock John Wiley \& Sons.

\bibitem[\protect\citeauthoryear{Howard and Matheson}{Howard and
  Matheson}{1981}]{Howard1981}
Howard, R. and J.~Matheson (1981).
\newblock {\em Influence diagrams}, Volume~II, pp.\  721--762.
\newblock Menlo Park, CA.

\bibitem[\protect\citeauthoryear{J{\"a}ggi, Schlegel, and Zanetti}{J{\"a}ggi
  et~al.}{2019}]{jaggi2019macroeconomic}
J{\"a}ggi, A., M.~Schlegel, and A.~Zanetti (2019).
\newblock Macroeconomic surprises, market environment, and safe-haven
  currencies.
\newblock {\em Swiss Journal of Economics and Statistics\/}~{\em 155\/}(1),
  1--21.

\bibitem[\protect\citeauthoryear{Johnston and Scott}{Johnston and
  Scott}{1999}]{johnston1999statistical}
Johnston, K. and E.~Scott (1999).
\newblock The statistical distribution of daily exchange rate price changes:
  dependent vs independent models.
\newblock {\em Journal of Financial and Strategic Decisions\/}~{\em 12\/}(2),
  39--49.

\bibitem[\protect\citeauthoryear{Kalisch, Hauser, Maathuis, and
  M{\"a}chler}{Kalisch et~al.}{2020}]{kalisch2020overview}
Kalisch, M., A.~Hauser, M.~Maathuis, and M.~M{\"a}chler (2020).
\newblock An overview of the pcalg package for r.

\bibitem[\protect\citeauthoryear{Karolyi}{Karolyi}{2004}]{karolyi2004does}
Karolyi, G.~A. (2004).
\newblock Does international financial contagion really exist?
\newblock {\em Journal of Applied Corporate Finance\/}~{\em 16\/}(2-3),
  136--146.

\bibitem[\protect\citeauthoryear{Kazemilari and Djauhari}{Kazemilari and
  Djauhari}{2013}]{kazemilari2013analysis}
Kazemilari, M. and M.~Djauhari (2013).
\newblock Analysis of a correlation network in world currency exchange market.
\newblock {\em Int. J. of Applied Mathematics and Statistics\/}~{\em 44\/}(14),
  202--209.

\bibitem[\protect\citeauthoryear{Keskin, Deviren, and Kocakaplan}{Keskin
  et~al.}{2011}]{keskin2011topology}
Keskin, M., B.~Deviren, and Y.~Kocakaplan (2011).
\newblock Topology of the correlation networks among major currencies using
  hierarchical structure methods.
\newblock {\em Physica A: Statistical Mechanics and its Applications\/}~{\em
  390\/}(4), 719--730.

\bibitem[\protect\citeauthoryear{Kwapie{\'n}, Gworek, Dro{\.z}d{\.z}, and
  G{\'o}rski}{Kwapie{\'n} et~al.}{2009}]{kwapien2009analysis}
Kwapie{\'n}, J., S.~Gworek, S.~Dro{\.z}d{\.z}, and A.~G{\'o}rski (2009).
\newblock Analysis of a network structure of the foreign currency exchange
  market.
\newblock {\em Journal of Economic Interaction and Coordination\/}~{\em
  4\/}(1), 55.

\bibitem[\protect\citeauthoryear{Laeven and Valencia}{Laeven and
  Valencia}{2018}]{laeven2018systemic}
Laeven, M.~L. and M.~F. Valencia (2018).
\newblock {\em Systemic banking crises revisited}.
\newblock International Monetary Fund.

\bibitem[\protect\citeauthoryear{Lancichinetti and Fortunato}{Lancichinetti and
  Fortunato}{2009}]{lancichinetti2009community}
Lancichinetti, A. and S.~Fortunato (2009).
\newblock Community detection algorithms: a comparative analysis.
\newblock {\em Physical review E\/}~{\em 80\/}(5), 056117.

\bibitem[\protect\citeauthoryear{Lauritzen}{Lauritzen}{2001}]{lauritzen2001causal}
Lauritzen, S.~L. (2001).
\newblock Causal inference from graphical models.
\newblock {\em Complex stochastic systems\/}, 63--107.

\bibitem[\protect\citeauthoryear{Maathuis, Kalisch, B{\"u}hlmann,
  et~al.}{Maathuis et~al.}{2009}]{maathuis2009estimating}
Maathuis, M.~H., M.~Kalisch, P.~B{\"u}hlmann, et~al. (2009).
\newblock Estimating high-dimensional intervention effects from observational
  data.
\newblock {\em The Annals of Statistics\/}~{\em 37\/}(6A), 3133--3164.

\bibitem[\protect\citeauthoryear{Mantegna}{Mantegna}{1999}]{mantegna1999hierarchical}
Mantegna, R.~N. (1999).
\newblock Hierarchical structure in financial markets.
\newblock {\em The European Physical Journal B-Condensed Matter and Complex
  Systems\/}~{\em 11\/}(1), 193--197.

\bibitem[\protect\citeauthoryear{Marcaccioli and Livan}{Marcaccioli and
  Livan}{2019}]{marcaccioli2019polya}
Marcaccioli, R. and G.~Livan (2019).
\newblock A p{\'o}lya urn approach to information filtering in complex
  networks.
\newblock {\em Nature communications\/}~{\em 10\/}(1), 1--10.

\bibitem[\protect\citeauthoryear{Meek}{Meek}{1995}]{Meek1995Uncertainty}
Meek, C. (1995).
\newblock Causal inference and causal explanation with background knowledge.
\newblock In P.~Besnard and S.~Hanks (Eds.), {\em Uncertainty in Artificial
  Intelligence}, Volume~11, pp.\  403–410.

\bibitem[\protect\citeauthoryear{Nier, Yang, Yorulmazer, and Alentorn}{Nier
  et~al.}{2007}]{nier2007network}
Nier, E., J.~Yang, T.~Yorulmazer, and A.~Alentorn (2007).
\newblock Network models and financial stability.
\newblock {\em Journal of Economic Dynamics and Control\/}~{\em 31\/}(6),
  2033--2060.

\bibitem[\protect\citeauthoryear{OECD}{OECD}{2021}]{OECD20210413}
OECD (2021).
\newblock Covid-19 spending helped to lift foreign aid to an all-time high in
  2020 but more effort needed.

\bibitem[\protect\citeauthoryear{Pearl}{Pearl}{1988}]{Pearl1988}
Pearl, J. (1988).
\newblock {\em ProbabiIistic Reasoning in Intelligent Systems: Networks of
  Plausible Inference}.
\newblock San Mateo, CA: MorganKanfmann.

\bibitem[\protect\citeauthoryear{Pearl}{Pearl}{2009}]{pearl2009causality}
Pearl, J. (2009).
\newblock {\em Causality}.
\newblock Cambridge university press.

\bibitem[\protect\citeauthoryear{Ranaldo and S{\"o}derlind}{Ranaldo and
  S{\"o}derlind}{2010}]{ranaldo2010safe}
Ranaldo, A. and P.~S{\"o}derlind (2010).
\newblock Safe haven currencies.
\newblock {\em Review of finance\/}~{\em 14\/}(3), 385--407.

\bibitem[\protect\citeauthoryear{Reuters}{Reuters}{2010}]{chinapost20100619}
Reuters (2010).
\newblock Timeline - china's reforms of the yuan exchange rate.

\bibitem[\protect\citeauthoryear{Rigobon}{Rigobon}{2019}]{rigobon2019contagion}
Rigobon, R. (2019).
\newblock Contagion, spillover, and interdependence.
\newblock {\em Econom{\'\i}a\/}~{\em 19\/}(2), 69--100.

\bibitem[\protect\citeauthoryear{Rodriguez}{Rodriguez}{2007}]{rodriguez2007measuring}
Rodriguez, J.~C. (2007).
\newblock Measuring financial contagion: A copula approach.
\newblock {\em Journal of empirical finance\/}~{\em 14\/}(3), 401--423.

\bibitem[\protect\citeauthoryear{Roubini}{Roubini}{2020}]{roubini2020coronavirus}
Roubini, N. (2020).
\newblock Coronavirus pandemic has delivered the fastest, deepest economic
  shock in history.
\newblock {\em The guardian\/}~{\em 25\/}(March).

\bibitem[\protect\citeauthoryear{Sarpong}{Sarpong}{2019}]{sarpong2019estimating}
Sarpong, S. (2019).
\newblock Estimating the probability distribution of the exchange rate between
  ghana cedi and american dollar.
\newblock {\em Journal of King Saud University-Science\/}~{\em 31\/}(2),
  177--183.

\bibitem[\protect\citeauthoryear{Scutari, Graafland, and Guti{\'e}rrez}{Scutari
  et~al.}{2018}]{scutari2018learns}
Scutari, M., C.~E. Graafland, and J.~M. Guti{\'e}rrez (2018).
\newblock Who learns better bayesian network structures: Constraint-based,
  score-based or hybrid algorithms?
\newblock In {\em International Conference on Probabilistic Graphical Models},
  pp.\  416--427. PMLR.

\bibitem[\protect\citeauthoryear{Serrano, Bogun{\'a}, and Vespignani}{Serrano
  et~al.}{2009}]{serrano2009extracting}
Serrano, M.~{\'A}., M.~Bogun{\'a}, and A.~Vespignani (2009).
\newblock Extracting the multiscale backbone of complex weighted networks.
\newblock {\em Proceedings of the national academy of sciences\/}~{\em
  106\/}(16), 6483--6488.

\bibitem[\protect\citeauthoryear{Soram{\"a}ki and Cook}{Soram{\"a}ki and
  Cook}{2016}]{cook2016network}
Soram{\"a}ki, K. and S.~Cook (2016).
\newblock {\em Network theory and financial risk}.
\newblock Risk Books.

\bibitem[\protect\citeauthoryear{Spirtes and Glymour}{Spirtes and
  Glymour}{1991}]{spirtes1991algorithm}
Spirtes, P. and C.~Glymour (1991).
\newblock An algorithm for fast recovery of sparse causal graphs.
\newblock {\em Social science computer review\/}~{\em 9\/}(1), 62--72.

\bibitem[\protect\citeauthoryear{Spirtes, Glymour, Scheines, and
  Heckerman}{Spirtes et~al.}{2000}]{spirtes2000causation}
Spirtes, P., C.~N. Glymour, R.~Scheines, and D.~Heckerman (2000).
\newblock {\em Causation, prediction, and search}.
\newblock MIT press.

\bibitem[\protect\citeauthoryear{Spirtes, Meek, and Richardson}{Spirtes
  et~al.}{1995}]{spirtes1995causal}
Spirtes, P., C.~Meek, and T.~Richardson (1995).
\newblock Causal inference in the presence of latent variables and selection
  bias.
\newblock In {\em Proceedings of the Eleventh conference on Uncertainty in
  artificial intelligence}, pp.\  499--506.

\bibitem[\protect\citeauthoryear{Stavroglou, Pantelous, Soramaki, and
  Zuev}{Stavroglou et~al.}{2017}]{stavroglou2017causality}
Stavroglou, S., A.~Pantelous, K.~Soramaki, and K.~Zuev (2017).
\newblock Causality networks of financial assets.
\newblock {\em The Journal of Network Theory in Finance\/}~{\em 3\/}(2),
  17--67.

\bibitem[\protect\citeauthoryear{Stiglitz}{Stiglitz}{2010}]{stiglitz2010contagion}
Stiglitz, J.~E. (2010).
\newblock Contagion, liberalization, and the optimal structure of
  globalization.
\newblock {\em Journal of Globalization and Development\/}~{\em 1\/}(2).

\bibitem[\protect\citeauthoryear{Sugihara, May, Ye, Hsieh, Deyle, Fogarty, and
  Munch}{Sugihara et~al.}{2012}]{sugihara2012detecting}
Sugihara, G., R.~May, H.~Ye, C.-h. Hsieh, E.~Deyle, M.~Fogarty, and S.~Munch
  (2012).
\newblock Detecting causality in complex ecosystems.
\newblock {\em science\/}~{\em 338\/}(6106), 496--500.

\bibitem[\protect\citeauthoryear{Su{\l}kowski et~al.}{Su{\l}kowski
  et~al.}{2020}]{sulkowski2020covid}
Su{\l}kowski, {\L}. et~al. (2020).
\newblock Covid-19 pandemic; recession, virtual revolution leading to
  de-globalization?
\newblock {\em Journal of Intercultural Management\/}~{\em 12\/}(1), 1--11.

\bibitem[\protect\citeauthoryear{Tumminello, Aste, Di~Matteo, and
  Mantegna}{Tumminello et~al.}{2005}]{tumminello2005tool}
Tumminello, M., T.~Aste, T.~Di~Matteo, and R.~N. Mantegna (2005).
\newblock A tool for filtering information in complex systems.
\newblock {\em Proceedings of the National Academy of Sciences\/}~{\em
  102\/}(30), 10421--10426.

\bibitem[\protect\citeauthoryear{Van~Rijckeghem and Weder}{Van~Rijckeghem and
  Weder}{2001}]{van2001sources}
Van~Rijckeghem, C. and B.~Weder (2001).
\newblock Sources of contagion: is it finance or trade?
\newblock {\em Journal of international Economics\/}~{\em 54\/}(2), 293--308.

\bibitem[\protect\citeauthoryear{Wang, Xie, Han, and Sun}{Wang
  et~al.}{2012}]{wang2012similarity}
Wang, G.-J., C.~Xie, F.~Han, and B.~Sun (2012).
\newblock Similarity measure and topology evolution of foreign exchange markets
  using dynamic time warping method: Evidence from minimal spanning tree.
\newblock {\em Physica A: Statistical Mechanics and its Applications\/}~{\em
  391\/}(16), 4136--4146.

\bibitem[\protect\citeauthoryear{Wang, Xie, Zhang, Han, and Chen}{Wang
  et~al.}{2014}]{wang2014dynamics}
Wang, G.-J., C.~Xie, P.~Zhang, F.~Han, and S.~Chen (2014).
\newblock Dynamics of foreign exchange networks: a time-varying copula
  approach.
\newblock {\em Discrete Dynamics in Nature and Society\/}~{\em 2014}.

\bibitem[\protect\citeauthoryear{Wen, Wei, and Huang}{Wen
  et~al.}{2012}]{wen2012measuring}
Wen, X., Y.~Wei, and D.~Huang (2012).
\newblock Measuring contagion between energy market and stock market during
  financial crisis: A copula approach.
\newblock {\em Energy economics\/}~{\em 34\/}(5), 1435--1446.

\bibitem[\protect\citeauthoryear{Yandell}{Yandell}{1997}]{yandell1997practical}
Yandell, B.~S. (1997).
\newblock {\em Practical data analysis for designed experiments}.
\newblock Routledge.

\end{thebibliography}
\newpage
\appendix
\section{PC Stable Algorithm}
\label{SecPC}
\paragraph{STEP 1 - Finding the Skeleton.}

We begin the procedure by linking all nodes of interest $X$ and link all of them to create a complete graph, as in Figure \ref{fig:complete}.

\noindent
\begin{minipage}[b]{.45\linewidth}
\begin{figure}[H]
\centering
\begin{tikzpicture}

\node[random] (xn) at (0,0) {$X_{N}$};
\node[random] (x4) at (4.5,1.5) {$X_{4}$};
\node[random] (x3) at (4.5,-1.5) {$X_{3}$};
\node[random] (x2) at (6.5,3) {$X_{2}$};
\node[random] (x1) at (6.5,-3){$X_{1}$};

\path[draw,thick,-]
(x1) edge (x2)
(x1) edge (x3)
(x1) edge (x4)
(x1) edge (xn)

(x2) edge (x3)
(x2) edge (x4)
(x2) edge (xn)
(x3) edge (x4)
(x3) edge (xn)
(x4) edge (xn);
\end{tikzpicture}
\captionof{figure}[A complete causal graph for the five financial instruments of interest.]{A complete causal graph for the five financial instruments of interest.}
\label{fig:complete}
\end{figure}
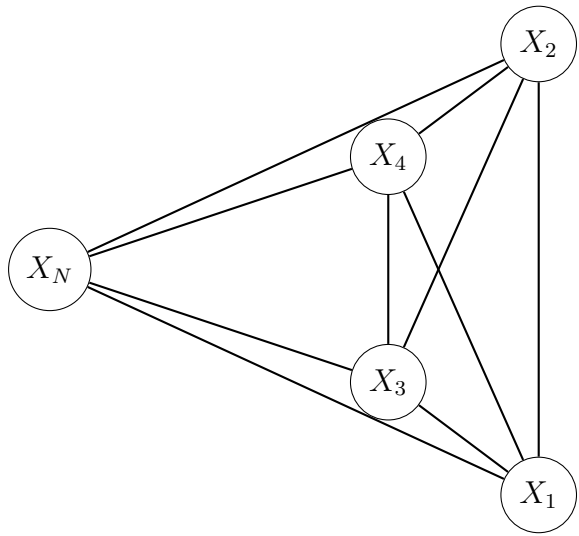
\end{minipage}%
\begin{minipage}[b]{.1\linewidth}
~
\end{minipage}%
\begin{minipage}[b]{.45\linewidth}
\begin{figure}[H]
\centering
\begin{tikzpicture}

\node[random] (xn) at (0,0) {$X_{N}$};
\node[random] (x4) at (4.5,1.5) {$X_{4}$};
\node[random] (x3) at (4.5,-1.5) {$X_{3}$};
\node[random] (x2) at (6.5,3) {$X_{2}$};
\node[random] (x1) at (6.5,-3){$X_{1}$};

\path[draw,thick,-]
(xn) edge (x4)
(xn) edge(x3)
(x4) edge (x3)
(x2) edge (x4)
(x3) edge (x1);
\end{tikzpicture}
\caption{The skeleton obtained from the complete graph using STEP 1.}
\label{fig:skeleton}
\end{figure}
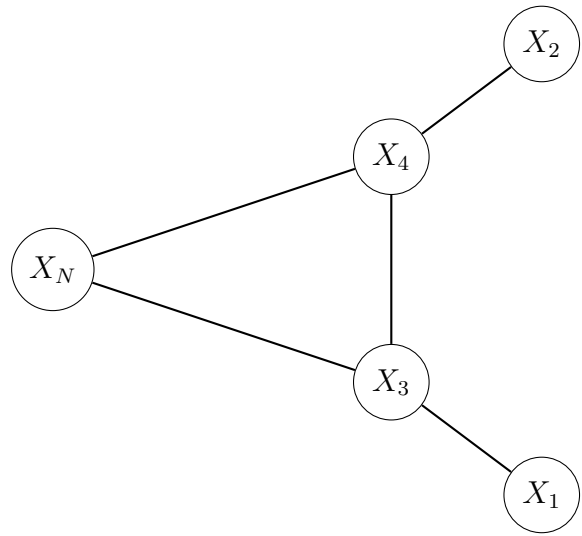
\end{minipage}

Once we have the complete graph, we eliminate links between nodes $X_{i}$ and $X_{j}$ that are independent or conditionally independent, conditioning iteratively on a growing subset of the other nodes $X_{C}$ in the graph. The algorithm considers all possible separating subsets $X_{C}$ for each pair of nodes to test the hypothesis $H_{0}$ that nodes $X_{i}$ and $X_{j}$ are independent. If there is any subset $X_{C}$ for which $H_{0}$ is not rejected at the specified significance level (often set at $\alpha=0.05$), the link between the two nodes is removed from the network. Whatever is left at the end of the process is the so-called skeleton, an undirected graph $G$ such that nodes $(X_{i},X_{j})$ are connected with a link if and only if no set $X_{C}$ can be found to make them conditionally independent. See an example of a skeleton in Figure \ref{fig:skeleton}.

\begin{figure}[tb]

\centering
\begin{tabular}{ccc|c}

\vspace{0pt}
\begin{tikzpicture}
\node[random] (xn) at (-1.5,1.5) {$X_{i}$};
\node[random] (x4) at (0,0) {$X_{k}$};
\node[random] (x2) at (1.5,1.5) {$X_{j}$};

\path[draw,thick,-]
(xn) edge (x4)
(x4) edge (x2);
\end{tikzpicture}
&
\vspace{0pt}
\begin{minipage}[c]{1cm}

$+$\\~~\\

\end{minipage}
&
\vspace{0pt}
\begin{minipage}[c]{3cm}

$X_{i} \indep X_{j}$\\~~\\

\end{minipage}
&
\vspace{0pt}
\begin{tikzpicture}

\node[random] (xn) at (-1.5,1.5) {$X_{i}$};
\node[random] (x4) at (0,0) {$X_{k}$};
\node[random] (x2) at (1.5,1.5) {$X_{j}$};

\path[draw,thick,color=red,->]
(xn) edge (x4)
(x2) edge (x4);
\end{tikzpicture}

\\ &{(A)} && (B)
\end{tabular}
\captionof{figure}[Example of a collider.]{Example of a collider. \par \small Knowing that $X_k$ is not a parent node for $X_i$ and/or $X_j$, only one direction for the links is logical possible.}
\label{fig:collider}
\end{figure}
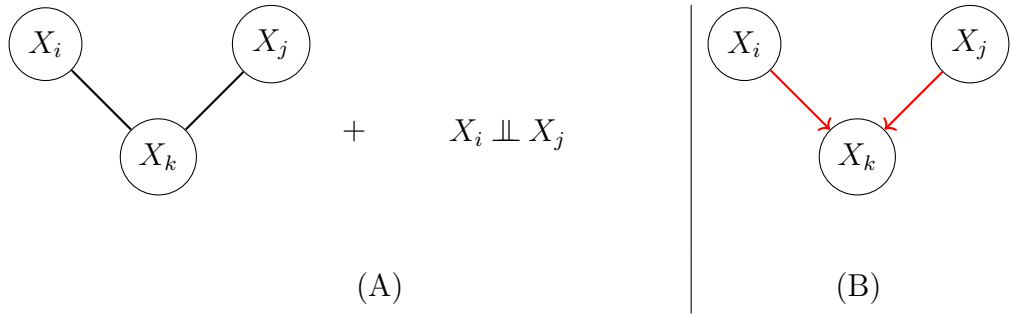

\paragraph{STEP 2 - Applying causal orientation rules.}
First, for each pair of nodes $X_{i}$ and $X_{j}$ that are not connected by a link, but both connecting to a common neighbour $X_{k}$, we check whether $X_{k}$ belongs to the subset of links $X_{C}$ from the previous step. Since we do not have a link between  $X_{i}$ and $X_{j}$, we know that a subset $X_{C}$ exists such that these two are conditionally independent. If $X_{k}$  does not belong to the subset $X_{C}$ and yet we still have a link between these three nodes, we know that $X_{i}$ and $X_{j}$ influence $X_{k}$ and not vice versa. This means we add directions $X_{i} \rightarrow X_{k}$ and $X_{j} \rightarrow X_{k}$ as seen in Figure \ref{fig:collider}. This primary orientation is often referred to as a collider or inverted fork or V orientation. We can think of colliders as nodes that stop a path unless the analysis is conditioned on them --- colliders are random variables that appear on the left-hand side in a structural equation model with multiple variables on the right-hand side.

%
%
%
%
%
%
%

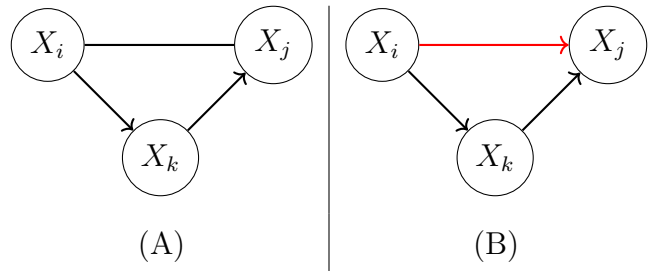
\begin{figure}[tb]

\centering
\begin{tabular}{c|c}

\begin{tikzpicture}
\node[random] (xn) at (-1.5,1.5) {$X_{i}$};
\node[random] (x4) at (0,0) {$X_{k}$};
\node[random] (x2) at (1.5,1.5) {$X_{j}$};

\path[draw,thick,-]
(xn) edge (x2);
\path[draw,thick,->]
(x4) edge (x2)
(xn) edge (x4);

\end{tikzpicture}
&
\begin{tikzpicture}
\node[random] (xn) at (-1.5,1.5) {$X_{i}$};
\node[random] (x4) at (0,0) {$X_{k}$};
\node[random] (x2) at (1.5,1.5) {$X_{j}$};

\path[draw,thick,color=red,->]
(xn) edge (x2);
\path[draw,thick,->]
(x4) edge (x2)
(xn) edge (x4);

\end{tikzpicture}

\\ (A) & (B)
\end{tabular}
\caption{Example of a consistency rule - avoiding a cycle.}
\label{fig:rule}
\end{figure}

%
%
%
%
Once we have established all colliders within the graph, the PC algorithm tries to orient as many of the remaining links as possible by a set of consistency rules, as in \cite{Meek1995Uncertainty}. The rules ensure that no newly directed link disrupts the previously established structure and no cycle is created. An example of such a rule is shown in Figure \ref{fig:rule}.

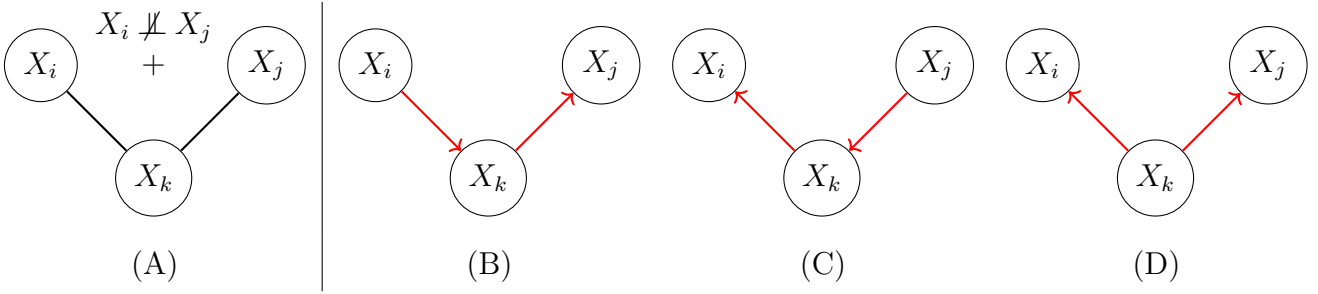
\begin{figure}[tb]

\centering
\begin{tabular}{c|ccc}
\begin{tikzpicture}

\node[random] (xn) at (-1.5,1.5) {$X_{i}$};
\node[random] (x4) at (0,0) {$X_{k}$};
\node[random] (x2) at (1.5,1.5) {$X_{j}$};
\node[] (x3) at (0,2) {$X_{i} \nindep X_{j}$};
\node[] (x3) at (0,1.5) {$+$};

\path[draw,thick,-]
(xn) edge (x4)
(x2) edge (x4);
\end{tikzpicture}
&
\begin{tikzpicture}
\node[random] (xn) at (-1.5,1.5) {$X_{i}$};
\node[random] (x4) at (0,0) {$X_{k}$};
\node[random] (x2) at (1.5,1.5) {$X_{j}$};

\path[draw,thick,color=red,->]
(xn) edge (x4)
(x4) edge (x2);
\end{tikzpicture}
&
\begin{tikzpicture}
\node[random] (xn) at (-1.5,1.5) {$X_{i}$};
\node[random] (x4) at (0,0) {$X_{k}$};
\node[random] (x2) at (1.5,1.5) {$X_{j}$};

\path[draw,thick,color=red,->]
(x4) edge (xn)
(x2) edge (x4);
\end{tikzpicture}
&
\begin{tikzpicture}
\node[random] (xn) at (-1.5,1.5) {$X_{i}$};
\node[random] (x4) at (0,0) {$X_{k}$};
\node[random] (x2) at (1.5,1.5) {$X_{j}$};

\path[draw,thick,color=red,->]
(x4) edge (xn)
(x4) edge (x2);
\end{tikzpicture}

\\ (A) & (B) & (C) & (D)
\end{tabular}
\captionof{figure}[Example of a CPDAG.]{Example of a CPDAG.  \par \small  (B),(C) and (D) are Markov equivalent DAGs, or CPDAGs, under the assumption that $X_k$ belongs to the subset of $X_C$ that made $X_i$ and $X_j$ independent and so a collider solution as in Figure \ref{fig:collider} was not possible.}
\label{fig:CPDAG}
\end{figure}

Not all links can always be successfully oriented at the end of this step. 
If the graph we obtain after implementing the orientation rules is not fully directed, it is referred to as a Completed Partially Directed Acyclic Graph (CPDAG). In that case a CPDAG is the best possible outcome we can obtain. It describes an equivalence class of DAGs that cannot be distinguished even with infinite amount of data.  An example of a CPDAG can be seen in Figure \ref{fig:CPDAG}. Unlike in Figure \ref{fig:collider}, we cannot find a collider, and so all three DAGs (B),(C) and (D) can equally explain the skeleton in (A).

The more nodes or variables we have in a network, and therefore the more connectivity we have, the easier it is for the PC Algorithm to detect directions. This is in  contrast to the  curse of dimensionality that most models have. A trivial example would be a network with just two nodes: In such a situation no amount of observational data on these two nodes would allow the PC Algorithm to orient the graph and decide on the direction of the arrow between the two nodes.

\end{document}